\documentclass[twocolumn,trackchanges]{aastex63}
\usepackage{hyperref}
\usepackage{amsmath}
\usepackage{graphicx}
\usepackage{soul}
\usepackage{kotex}

\newcommand{\ramses}{\texttt{RAMSES}}
\newcommand{\music}{\texttt{MUSIC}}

\newcommand{\msun}{M_{\odot}}
\newcommand{\ysamtm}{\texttt{ySAMtm}}
\newcommand{\pgalf}{\texttt{PGalF}}

\newcommand{\hr}{\texttt{HR5}}
\newcommand{\hrl}{\texttt{HR5-Low}}

\newcommand{\rturn}{$R_{\rm TA}$}

\definecolor{Orange}{rgb}{1.0,0.5,0.15}
\definecolor{Blue}{rgb}{0,0.08,0.65}
\definecolor{Blue2}{rgb}{0,0.4,0.6}

\definecolor{Red}{rgb}{0.65,0.08,0.05}
\definecolor{Green}{rgb}{0.15,0.45,0.25}
\definecolor{Pink}{rgb}{1.0,0.05,0.5}
\definecolor{Purple}{rgb}{0.3,0.,0.5}

\definecolor{red}{RGB}{250,0,0}
\definecolor{blue}{RGB}{0,0,250}

\usepackage{soul} 
\usepackage{xcolor}
\definecolor{aquamarine}{rgb}{0.5, 1.0, 0.83}

\definecolor{RedWine}{rgb}{0.743,0,0}

\shorttitle{Identification of Protoclusters}
\shortauthors{Lee et al.}

\begin{document}
\title{Identification of Galaxy Protoclusters Based on the Spherical Top-hat Collapse Theory}

\author[0000-0002-6810-1778]{Jaehyun Lee}
\affiliation{Korea Astronomy and Space Science Institute, 776 Daedeokdae-ro, Yuseong-gu, Daejeon 34055, Korea}
\affiliation{Korea Institute for Advanced Study, 85 Hoegi-ro, Dongdaemun-gu, Seoul 02455, Korea}

\author[0000-0001-9521-6397]{Changbom Park}
\affiliation{Korea Institute for Advanced Study, 85 Hoegi-ro, Dongdaemun-gu, Seoul 02455, Korea}

\author[0000-0002-4391-2275]{Juhan Kim}
\affiliation{Center for Advanced Computation, Korea Institute for Advanced Study, 85 Hoegiro, Dongdaemun-gu, Seoul 02455, Korea}

\author[0000-0003-0695-6735]{Christophe Pichon}
\affiliation{Institut d'Astrophysique de Paris, CNRS and Sorbonne Universit\'e, UMR 7095, 98 bis Boulevard Arago, F-75014 Paris, France}
\affiliation{ IPhT, DRF-INP, UMR 3680, CEA, L'Orme des Merisiers, B\^at 774, 91191 Gif-sur-Yvette, France}
\affiliation{Korea Institute for Advanced Study, 85 Hoegi-ro, Dongdaemun-gu, Seoul 02455, Korea}

\author[0000-0003-4446-3130]{Brad K. Gibson}
\affiliation{E.A. Milne Centre for Astrophysics, University of Hull, Hull, HU6 7RX, United Kingdom}

\author[0000-0001-5135-1693]{Jihye Shin}
\affiliation{Korea Astronomy and Space Science Institute, 776 Daedeokdae-ro, Yuseong-gu, Daejeon 34055, Korea}

\author[0000-0003-4164-5414]{Yonghwi Kim}
\affiliation{Korea Institute of Science and Technology Information, 245 Daehak-ro, Yuseong-gu, Daejeon, 34141, Korea}

\author[0000-0003-1414-1296]{Owain N. Snaith}
\affiliation{University of Exeter, School of Physics and Astronomy, Stocker Road, Exeter, EX4 4QL, UK}

\author[0000-0003-0225-6387]{Yohan Dubois}
\affiliation{CNRS and Sorbonne Universit\'e, UMR 7095, Institut d'Astrophysique de Paris, 98 bis, Boulevard Arago, F-75014 Paris, France}

\author{C. Gareth Few}
\affiliation{E.A. Milne Centre for Astrophysics, University of Hull, Hull, HU6 7RX, United Kingdom}

\correspondingauthor{Juhan Kim}
\email{kjhan@kias.re.kr}

\begin{abstract}

We propose a new method for finding galaxy protoclusters that is motivated by structure formation theory and also directly applicable to observations. 
We adopt the conventional definition  that a protocluster is a galaxy group whose virial mass $M_{\rm vir} < M_{\rm cl}$ at its epoch, where $M_{\rm cl}=10^{14}\,\msun$, but would exceed that limit when it evolves to $z=0$. We use the critical overdensity for complete collapse at $z = 0$ predicted by the spherical top-hat collapse model to find the radius and total mass of the regions that would collapse at $z=0$. If the mass of a region centered at a massive galaxy exceeds $M_{\rm cl}$, the galaxy is at the center of a protocluster. We define the outer boundary of protocluster as the zero-velocity surface at the turnaround radius so that the member galaxies are those sharing the same protocluster environment and showing some conformity in physical properties. We use the cosmological hydrodynamical simulation Horizon Run 5 (\hr) to calibrate this prescription and demonstrate its performance. We find that the protocluster identification method suggested in this study is quite successful. Its application to the high redshift \hr\ galaxies shows a tight correlation between the mass within the protocluster regions identified according to the spherical collapse model and the final mass to be found within the clusters at $z=0$, meaning that the regions can be regarded as the bona-fide protoclusters with high reliability. We also confirm that the redshift-space distortion does not significantly affect the performance of the protocluster identification scheme.

\end{abstract}





\section {Introduction}
\label{sec:intro}



Galaxy clusters are typically defined as the objects that are bound and dynamically relaxed with total mass of $M_{\rm tot}>10^{14}\,\msun$~\citep[e.g.,][]{overzier16}. As the progenitors of present-day galaxy clusters, protoclusters must have formed in the densest environments in the early universe, and the majority of the galaxies in protoclusters probably have formed and evolved earlier than those in other environment~\citep{Kaiser1984}. 

Many observational efforts have been made to search for protoclusters at high redshifts.
Deep-field spectroscopic survey is a direct approach for finding 
protoclusters~\citep[e.g.,][]{steidel98,steidel00,steidel05,lee14,toshikawa14,cucciati14,lemaux14,chiang15,diener15,wang16,calvi21,mcconachie22}. 
However, the survey volume should be very large to include many of such rare objects, and spectroscopic observations are currently too time-consuming to carry out large-volume blind surveys for the deep universe. 
Therefore, large-area imaging surveys have been often conducted to search for overdense regions at high redshifts by utilizing the narrow-band photometry for emission-line galaxies or the photo-$z$/dropout technique
\citep[e.g.,][]{shimasaku03,ouchi05,toshikawa12,toshikawa16,cai17a,toshikawa18,shi19,yonekura22}. 

Some energetic events are expected to happen in overdense regions at high redshifts. High-$z$ radio galaxies are believed to be the potential progenitors of brightest cluster galaxies and, thus, they are assumed as a proxy for protoclusters~\citep{pascarelle96,lefevre96,venemans02,venemans04,venemans05,venemans07,hatch11a,hatch11b,hayashi12,cooke14,shen21}. Although it is still  debated~\citep[see][]{husband13,hennawi15}, high-$z$ QSOs are also known to trace overdense regions~\citep{djorgovski03,wold03,stevens10,falder11,adams15}. Ly$\alpha$ blobs can be lit by a huge amount of ionized photons emitted from AGNs or starburst galaxies in dense regions which still bear sufficient cold gas as a fuel. High-$z$ submillimeter galaxies 
are regarded as the progenitors of massive ellipticals~\citep[e.g.,][]{lilly99,fu13,toft14}. Therefore, Ly$\alpha$ blobs or overdensity regions of submillimeter galaxies are also used as the indicators of protocluster regions~\citep{stevens03,greve07,prescott08,daddi09,prescott12,umehata14,umehata15,oteo18,cooke19,rotermund21,alvarezcrespo21}. Gas absorption lines are another probe of protoclusters that does not rely on galaxy distribution: high-$z$ overdense regions that still contain plenty of intergalactic neutral hydrogen can be detected by examining the Ly$\alpha$ forests in the spectra of background QSOs or star-forming galaxies~\citep[e.g.,][]{lee14a,stark15,cai16,cai17a,newman22}.

While the observations targeting protoclusters have used a variety of selection techniques, they commonly focus on the identification of overdense regions. 
The protoclusters that are expected to eventually form massive clusters with the total mass of $M_{\rm tot}>10^{15}\,\msun$ have the overdensity of $\delta\sim10-12$ for typical galaxies or Ly$\alpha$ emitters within an aperture radius of $R\sim15\,$cMpc at $z\sim2-3$~\citep[e.g.,][]{lemaux14,cucciati14,cai17a}. \citet{toshikawa18} identify protocluster candidates in a wide field of $>100\,$deg$^2$ by selecting the regions that show the galaxy overdensity significance level higher than $4\sigma$ within an aperture radius of $R\sim16\,$cMpc at $z\sim3.8$. This significance level corresponds to the overdensity of the regions ending up forming halos of $M_{\rm halo}\gtrsim5\times10^{14}\,\msun$. The overdensity significance level is adopted to achieve $\sim80\%$ reliability, at the cost of completeness~\citep{toshikawa16}

Several theoretical studies have been conducted to examine the properties of protocluster regions. \citet{chiang13} and \citet{muldrew15} investigate the matter and galaxy overdensity in the areas enclosing protoclusters using the semi-analytic model of \citet{guo11} based on the Millennium simulation~\citep{springel05}. In the two studies, protoclusters are traced using halo merger trees. They show that the protocluster galaxies are more widespread in larger clusters, and the distribution of protocluter galaxies 
largely shrink during $z=4-2$. \citet{chiang13} also show that, in a top-hat box of $\rm (15\,cMpc)^3$, the galaxy overdensity of protoclusters strongly correlates with final cluster mass. \citet{wang21} develop a method to identify protoclusters from halo distribution of an N-body simulation using an extension of the Friend-of-Friend (FoF) algorithm. They show that the approach reasonably recovers protoclusters with high completeness.

Hydrodynamical simulations are also used to study the formation and evolution of  clusters of galaxies. Given that the mean separation of rich clusters is $\sim70\,$cMpc~\citep{bachcall92},   it is thus necessary to use a simulation box larger than  about $1\,$cGpc$^3$ 
to study the formation and evolution of Coma-like clusters accurately and with high statistical significance. However, due to the limitation of the current computing resources, it has been nearly impossible to conduct hydrodynamical simulations in such a large box while keeping a resolution below $\sim 1$kpc. As a compromise between the need for the extremely large dynamic range and the limited computing resources, the zoom-in technique is widely adopted in the hydrodynamical simulations for galaxy clusters~\citep{bahe17,choi17,truong18,yajima22,Trebitsch2021}. In these simulations, cluster regions are pre-identified and zoomed in the initial conditions, and 
protoclusters are traced by using merger trees.

It should be noted that, in the previous studies, protoclusters have been 
defined inconsistently between observations, theories, and numerical simulations. If a protocluster is defined as the group of all the objects that will eventually collapse into a cluster, their initial distribution typically spans more than tens of cMpc~\citep{chiang13,muldrew15,muldrew18}. In this definition, protoclusters can be neither self-bound nor compact, and thus a protocluster is hardly viewed as a physical object in which galaxies are associated with each other in a common environment. Furthermore, diachronic information is not available in observations. Therefore, observers have focused on the identification of sufficiently overdense regions. This is justified by the fact that larger structures in the current universe are more likely to originate from more massive progenitors at high redshifts~\citep{chiang13,muldrew15}. The range of overdense region varies between protoclusters. Since the virial radius only encloses the objects which are already bound to the local density peak, it inevitably misses a number of progenitors which are still in the course of infall, outside the virialized regions. Because the proto-objects of larger clusters are more extended~\citep{muldrew15}, a systematic approach is required to define the boundary (or spatial extent) of protoclusters, which should be based on the physical conditions of specific environments of interest.

This study aims at proposing a new scheme for the identification of protoclusters that is motivated by structure formation theories, and also applicable to observations directly. Our prescription is justified and calibrated on  
a cosmological hydrodynamical simulation Horizon Run 5~\citep[hereafter \hr,][]{lee21,park22}.
\hr\ covers a volume of (1048.6~cMpc)$^3$ with a spatial resolution down to about 1~kpc. 
Thanks to its large volume, \hr\ enables us to look into the formation and evolution of galaxies in a wide range of environments. 
By taking advantage of \hr, we derive a scheme applicable to observations to find the centers of protocluster candidates based on the spherical top-hat collapse (SC) model. The scheme also defines the physical region of a given protocluster as the volume within the turnaround radius from their centers. The turnaround radius is the zero-velocity surface at which gravitational infall counterbalances the local Hubble expansion~\citep{gunn72}.


This paper is organized as follows. In Section~\ref{sec:simulation}, we briefly introduce the \hr\ simulation, a structure finding and a tree building algorithm, and the scheme to identify clusters using a low resolution version of \hr. In Section~\ref{sec:protoclusters}, we present the methodology to find the candidate regions for protoclusters from the galaxy distribution.
The method for finding the boundary of protoclusters is presented
in Section~\ref{sec:turnaround_radius}. We discuss and summarize this study in Section~\ref{sec:summary}. Additional details of structure identification, merger tree building schemes, the SC models, and protocluster identification are given in Appendix.

\section{Simulation Data}
\label{sec:simulation}

\subsection{Horizon Run 5}
\hr\ is a cosmological hydrodynamical zoomed simulation aiming at covering a wide range of cosmic structures in a 1.15\,cGpc$^3$ volume, with a spatial resolution down to $\sim1\,$kpc. We adopt the cosmological parameters of $\Omega_{\rm m}=0.3$, $\Omega_{\Lambda}=0.7$, $\Omega_{\rm b}=0.047$, $\sigma_8=0.816$, and $h=0.684$ that are compatible with the Planck data~\citep{planck16}. 
We generate the initial conditions using the \music\ package~\citep{hahn11}, with a second-order Lagrangian scheme to launch the particles~\citep[2LPT;][]{scoccimarro98,lhuillier14}. \hr\ is conducted using a version of the adaptive mesh refinement code \ramses~\citep{teyssier02} upgraded for an OpenMP plus MPI two-dimensional parallelism~\citep[]{lee21}. We generated a number of random sets and selected the one that reproduced the theoretical baryonic acoustic oscillation features most closely. While the volume of the zoomed region is still somewhat insufficient for accurate statistical analyses of the most massive galaxy clusters and the impact of the very large-scale structures, the whole simulation box does manage to encompass the relevant large-scale perturbation modes, and provides us with a representative volume corresponding to the input cosmology.

\begin{figure*}
\centering 
\includegraphics[width=0.95\textwidth]{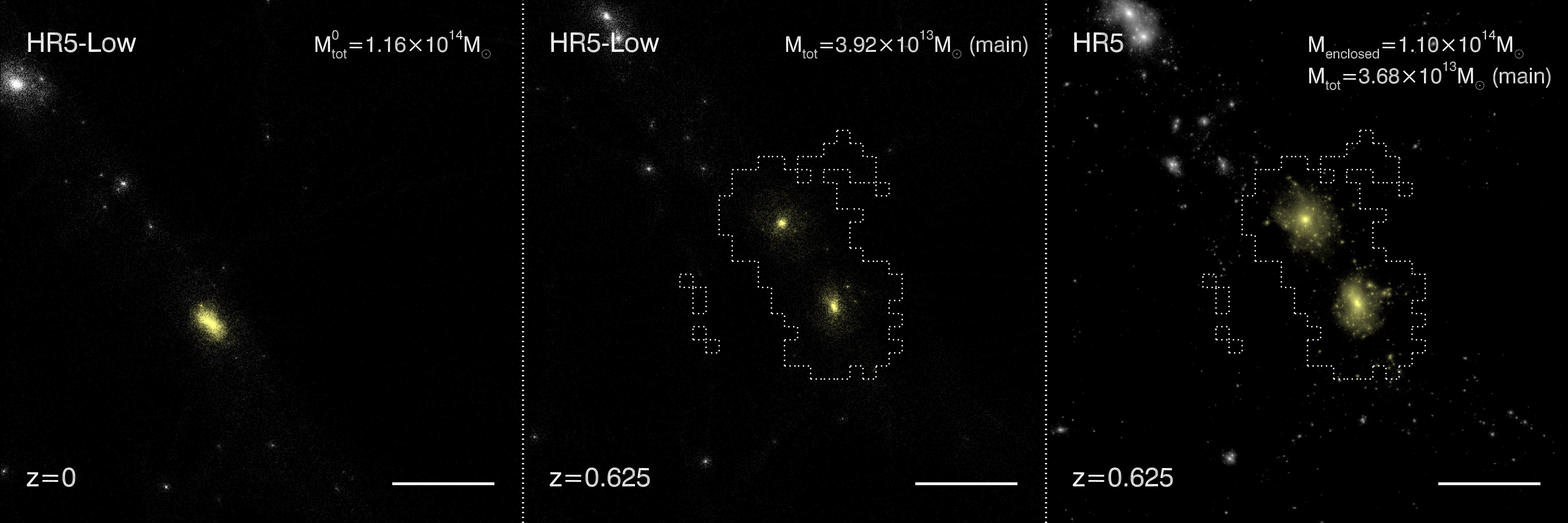}
\includegraphics[width=0.95\textwidth]{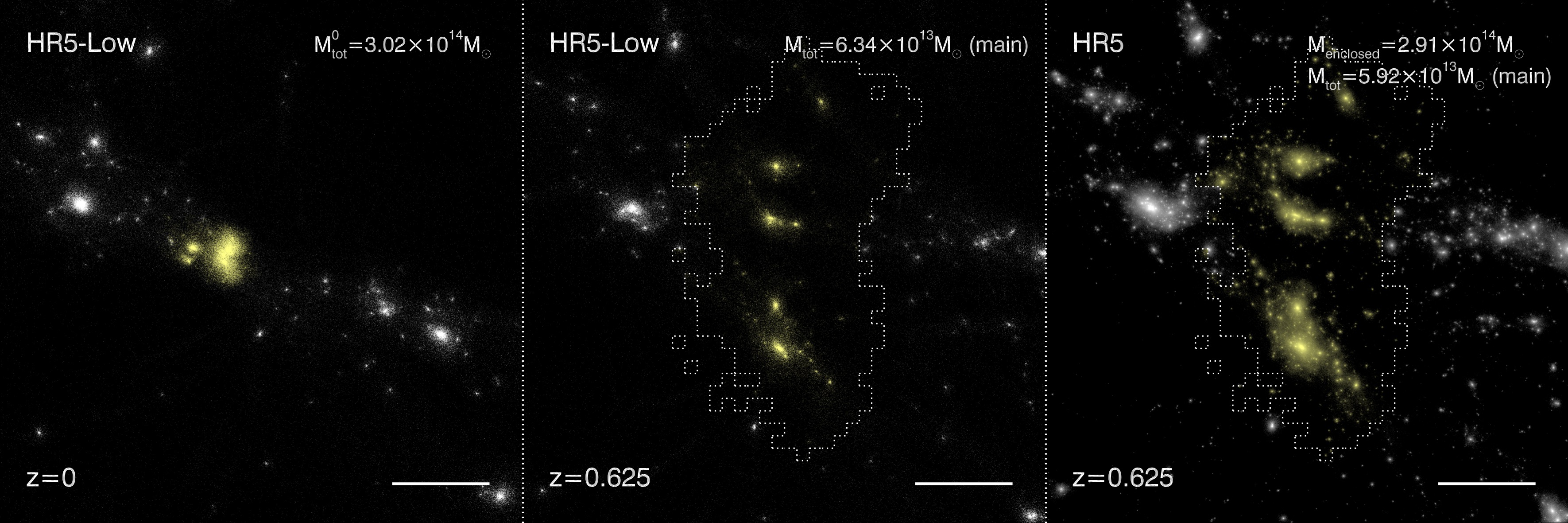}
\includegraphics[width=0.95\textwidth]{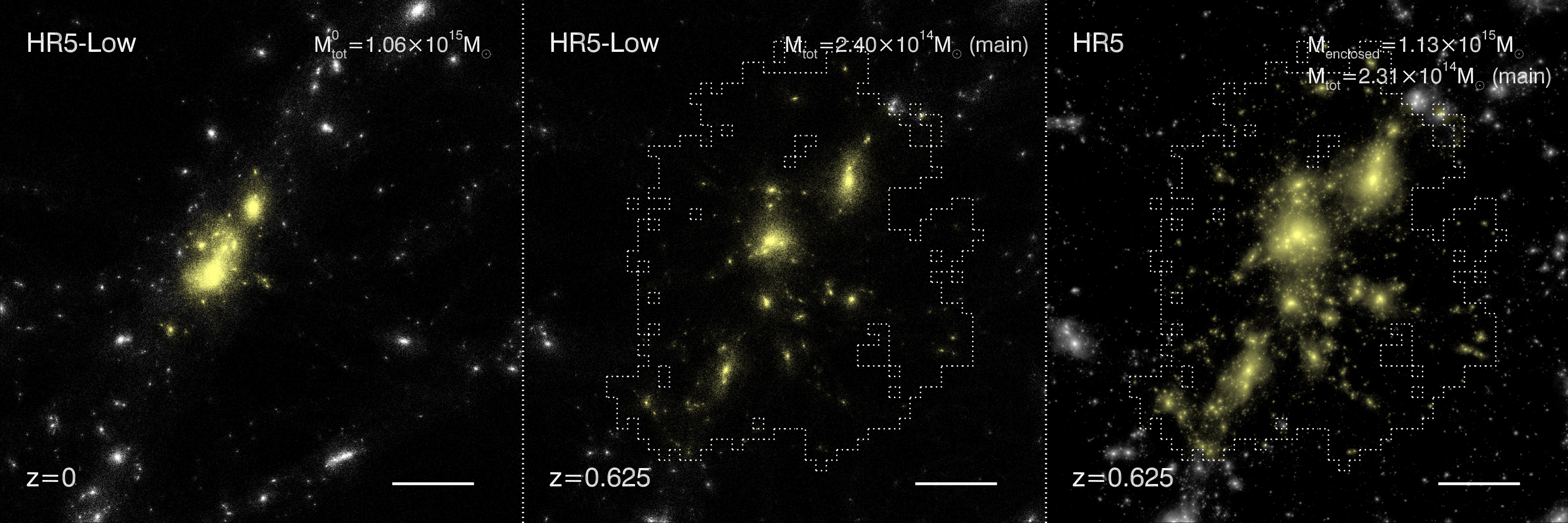}

\caption{Dark matter particles in three clusters found at $z=0$ in \hrl\ (left), in their progenitors at $z=0.625$ (middle), and in the same volumes in \hr\ at $z=0.625$ (right). In the left panels, the halos in yellow are the members of the clusters with $M^0_{\rm tot}\sim10^{14}\,\msun$ (top), $10^{14.5}\,\msun$ (middle), and $10^{15}\,\msun$ (bottom). White horizontal bars illustrate the scale of 4cMpc. 
The white dotted lines display the Lagrangian volumes enclosing the dark matter particles that end up forming clusters at $z=0$ in \hrl. We assume that all the objects in \hr\ located inside the same Lagrangian volume are the progenitors of the corresponding cluster. The thickness of the projected volume is 8.2\,cMpc (top), 13.8\,cMpc (middel), and 21.5\,cMpc (bottom), fully containing each cluster in the projected direction. In the right panels, $M_{\rm enclosed}$ presents the total mass enclosed by the Lagrangian volume. 
All the objects inside the Lagrangian volume are traced back to high redshifts using their merger trees in this study.}
\label{fig:dm_proj_comp}
\end{figure*}

The volume of \hr\ is set to have a high-resolution cuboid zoomed region of $1048.6\times119.0\times127.2\,{\rm cMpc}^3$ crossing the center of the volume. The effective volume of the region is $\sim(260\,{\rm cMpc})^3$. The cosmological box has 256 root cells (level 8, $\Delta x=4.10\,$cMpc) on a side and the zoomed region has 8192 cells (level 13, $\Delta x=0.128\,$cMpc) along the long side in the initial conditions. The high-resolution region initially contains $8192\times930\times994$ cells and dark matter particles, and is surrounded by the padding grids of levels from 12 to 9. The dark matter particle mass is $6.89\times10^7\,\msun$ in the zoomed region, and increases by a factor of 8 with a decreasing grid level. The cells are adaptively refined down to $\Delta x\sim1\,$kpc when their density exceeds eight times the dark matter particle mass at level 13. \hr\ was proceeded through $z=0.625$.

Physical processes driving the evolution of baryonic components are implemented in subgrid forms in \ramses. Gas cooling is computed using the cooling functions of \citet{sutherland93} in a temperature range of $10^4-10^{8.5}\,$K and fine-structure line cooling is computed down to $\sim750\,$K using the cooling rates of \citet{dalgarno72}. \ramses\ approximates cosmic reionization by assuming a uniform UV background~\citep{haardt96}. The statistical approach of \citet{rasera06} is adopted to compute a star formation rate. Supernova feedback affects the interstellar medium in thermal and kinetic modes~\citep{dubois08} and AGN feedback operates in radio-jet and quasar modes, relying on the Eddington ratio~\citep{dubois12}. Massive black holes (MHBs) are seeded with an initial mass of $10^4\,\msun$ in grids when gas density is higher than the threshold of star formation and no other MBHs is found within 50 kpc~\citep{dubois14b}. MBHs grow via accretion and coalescence, and their angular momentum obtained from the feeding processes are traced~\citep{dubois14a}. Metal enrichment is computed using the method proposed by \citet{few12} based on a Chabrier initial mass function~\citep{chabrier03}, and in particular the abundance of H, O, and Fe are traced individually. One can find further details of \hr\ in \citet{lee21}.

\begin{figure*}
\centering 
\includegraphics[width=0.8\textwidth]{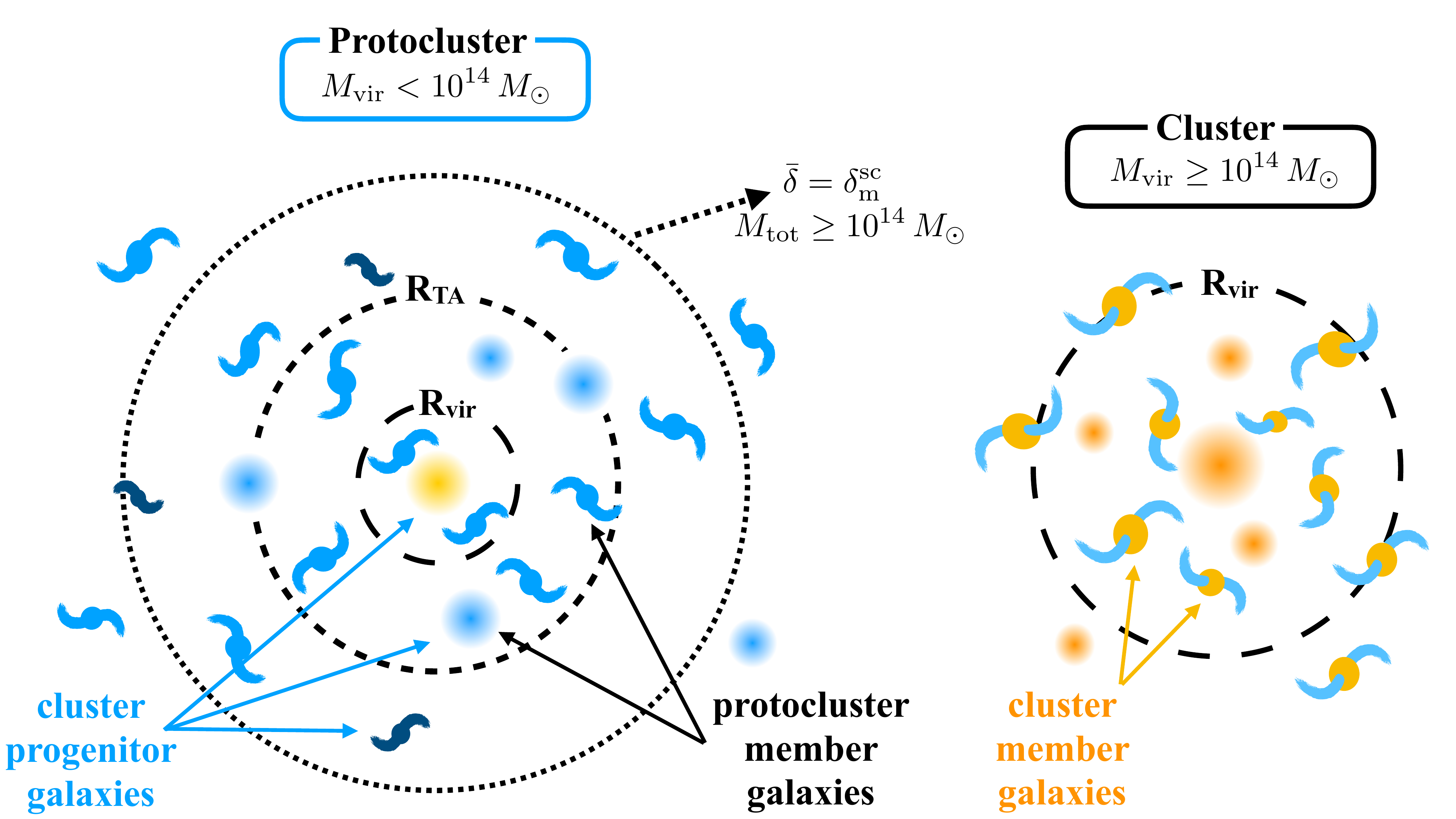}
\caption{A schematic diagram presenting the definition of galaxy protoclusters and clusters. Clusters are groups of galaxies with $M_{\rm vir}$ currently greater than $ 10^{14}\msun$. Protoclusters are those with $M_{\rm vir} < 10^{14}\msun$ currently, but will have $M_{\rm vir} \ge 10^{14}\msun$ by $z=0$. The future virial mass is estimated from the total mass within the region having the mean overdensity $\bar{\delta}$ equal to the critical overdensity $\delta_{\rm m}^{\rm sc}$ for complete collapse at $z=0$ predicted by the spherical top-hat theory. The physical volume of protoclusters is defined to be the region within the turn-around radius $R_{\rm TA}$.}
\label{fig:definition}
\end{figure*}


\subsection{Identification of Clusters Using a Low Resolution Simulation}
\label{sec:identification_of_cluster}

We identify FoF halos and self-bound objects embedded in FoF halos using \pgalf~\citep{kim+22}. We also construct the merger trees of self-bound objects using \ysamtm~\citep{jung14,leej14} based on stellar particles for galaxies and dark matter particles for halos that contain no stars. The details of the structure finding and tree building algorithms are given in \ref{sec:merger_trees}. 

In this study, we define a galaxy cluster as the virialized object that has acquired the total mass of $M_{\rm tot}>10^{14}\,\msun$ at or before $z=0$. The mass cut is adopted following the conventional mass range of galaxy clusters~\citep[e.g.,][]{overzier16}, and can be varied if a different mass range is necessary. Protoclusters are the progenitors of galaxy clusters that have not reached the cluster-scale mass range yet. By this definition, both clusters and protoclusters can be found at any epoch. 
According to this definition of a galaxy cluster, we cannot directly identify all the clusters and protoclusters in \hr \ as the simulation stopped at $z=0.625$. 
At this redshift, we find 63 clusters with $M_{\rm tot}>10^{14}\,\msun$ in the zoomed region. Objects having mass contamination higher than 0.7\% by the lower level particles are excluded. 
However, there can be many structures that are not massive enough to be identified as clusters at $z=0.625$ but will evolve to cluster-scale halos by $z=0$. 

\begin{figure*}
\centering 
\includegraphics[width=0.95\textwidth]{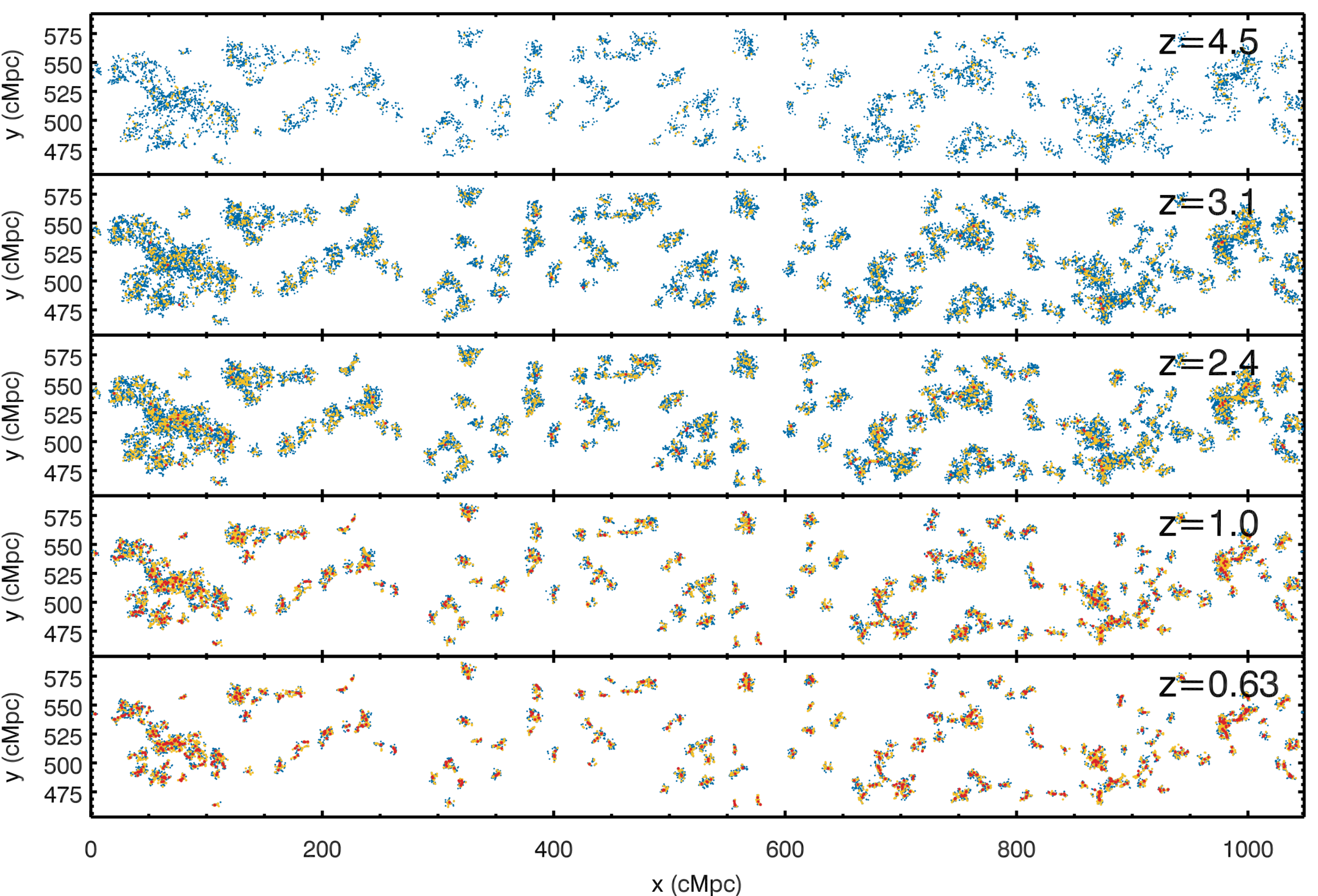}
\caption{Distribution of cluster progenitor galaxies in \hr\ at $z=0.63-4.5$. Blue, yellow and red dots mark the locations of the progenitor galaxies with stellar mass of $\log M_\star/\msun=9-10$, $10-11$, and $>11$, respectively.}
\label{fig:proto_distribution}
\end{figure*}

To find clusters and protoclusters in the last snapshot of \hr\ (i.e. $z=0.625$), we additionally conduct a low-resolution simulation \hrl\,($\Delta x\sim16\,$kpc) based on the initial conditions and the model parameters used in \hr. We identify structures from the snapshots of \hrl\ at $z=0$ and 0.625 using \pgalf. At $z=0$, we find 2,794 objects of $M^0_{\rm tot}\ge10^{13}\,\msun$ and 189 objects of $M^0_{\rm tot}\ge10^{14}\,\msun$ with the contamination tolerance mentioned above. The dark matter particles are traced back to $z=0.625$ using their IDs, to search for the progenitors of the clusters. We then construct the Lagrangian volume~\citep[hereafer LV, for details see][]{onorbe+14} of the progenitors using the uniform cubic grids enclosing the dark matter particles finally assembling the clusters. We assume that the LVs constructed from \hrl\ also enclose the clusters or protoclusters in \hr. We present the details of the identification scheme and reliability of this approach in \ref{sec:cluster_candidates}. Figure~\ref{fig:dm_proj_comp} shows the dark matter distribution in three \hrl\ cluster regions at $z=0$ ({\it left}), the same regions of \hrl\ ({\it middle}), and \hr\ ({\it right}) at $z=0.625$. The structure colored in yellow is the FoF halo of each cluster ({\it left}), its progenitors at $z=0.625$ ({\it middle}), and its counterpart in \hr\ ({\it right panels}). The grids enclosed by dotted lines mark the LVs of the objects constructed by tracing the dark matter particles.
This figure demonstrates that the two simulations are in good agreement despite their different resolutions. The position of a structure may show a slight offset between the two different resolution simulations (\hrl\ and \hr) at $z=0.625$ partly due to the adaptive time step in \ramses.


\section{Identification of protoclusters}
\label{sec:protoclusters}

We define `protoclusters' as galaxy groups whose total mass within $R_{\rm vir}$ is currently less than $10^{14}\msun$ at their epochs but would exceed that limit by $z=0$. 
The physical extent of a protocluster is defined as the spherical volume within the turnaround radius or the zero-velocity surface. The concept is schematically visualized in Figure~\ref{fig:definition}.
A protocluster is located at the center of a sphere that has the mean density ${\bar \delta}$ and encloses the total mass exceeding $10^{14} \msun$. The critical overdensity $\delta^{\rm sc}_{\rm m} = {\bar \delta}$ is given by the spherical top-hat theory, and the mass contained is the expected virial mass of the region at $z=0$. It should be noted that only the galaxies within the turnaround radius are called the protocluster member galaxies, and that the cluster progenitor galaxies can be spread out to much larger radii.

We first identify the authentic proto-objects by tracing their merger trees in Section~\ref{sec:merger_trees_proto}, and then present a systematic approach for finding the candidate regions enclosing protoclusters from the galaxy distribution in a snapshot, without diachronic information, in Section~\ref{sec:sc_model_proto}.

\subsection{Identification of Proto-objects using Merger Trees}
\label{sec:merger_trees_proto}

We search for the bona-fide progenitors of each cluster or protocluster of \hr\ at $z=0.625$ by tracing backward their merger histories. 
All the progenitors of each object are identified in all snapshots. Note that we do not call all the progenitors the protocluster galaxies, as protocluster galaxies will be defined as those within the turnaround radius.
We define the most massive galaxy among the progenitors in a snapshot as the central galaxy. Thus, the central galaxy of a protocluster may change over time, depending on their mass accretion history.

The bottom panel of Figure~\ref{fig:proto_distribution} shows the distribution of the galaxies belonging to clusters or protoclusters in comoving space at $z=0.625$.
The upper four panels show their progenitors that are traced along merger trees. Red, yellow, and blue dots mark the galaxies with $M_\star>10^{11}\,\msun$, $10^{10}-10^{11}\,\msun$, and $10^{9}-10^{10}\,\msun$, respectively. 
It can be seen that the overall locations of protoclusters hardly change over time: the initial conditions are essentially preserved for these massive objects sitting at deep gravitational potential minima.
On the other hand, the systems of cluster progenitor galaxies have been monotonically shrinking since $z\sim 2.4$. However, at redshifts higher than $z\sim 2.4$, their extent is roughly static at the value of
$R\sim10-30$\,cMpc, and the systems start to fade away.
The three redshifts of $z=2.4$, 3.1, and 4.5 are the target redshifts of the ODIN survey for LAEs at $z=2.4$, 3.1, and 4.5 \citep{ramakrishnan22}. We will discuss the results of this study mainly at these redshifts. 

\subsection{Identification of Protocluster Candidates based on the Spherical Top-Hat Collapse Model}
\label{sec:sc_model_proto}

In this subsection, we propose a systematic method to identify the candidate regions enclosing protoclusters from galaxy distribution based on the SC models.

\subsubsection{Overdensity threshold for complete collapse at $z=0$}
\label{sec:sc_overdensity}

We define protocluster candidate regions as the spherical volumes that enclose total mass greater than $10^{14}\,\msun$ and will collapse completely at $z=0$ according to the overdensity threshold given by the spherical top-hat collapse model. We will search for the centers of protoclusters inside the spherical regions.

In the spherical top-hat collapse model, an overdense region at an epoch will contract into a point at some stage if its overdensity is equal to the critical threshold density. We find 
this threshold density as a function of redshift for two types of cosmology.
In the Einstein de-Sitter (EdS) universe with $\Omega_{\rm m}=1$,  a homogeneous density sphere that collapses at $z=0$ reaches its maximum radius at $z=0.59$ with $\delta_{\rm m}^{\rm sc}=9\pi^2 /16 -1\simeq 4.55$, where $\delta_{\rm m}^{\rm sc}$ is the spherical top-hat matter overdensity. See \ref{sec:EdS} for more details. For comparison, the linear theory predicts overdensity $\delta_{\rm m}^{\rm lin}\simeq1.062$ at $t_{\rm max}$ in the EdS universe. 

\begin{figure}
\centering 
\includegraphics[width=0.45\textwidth]{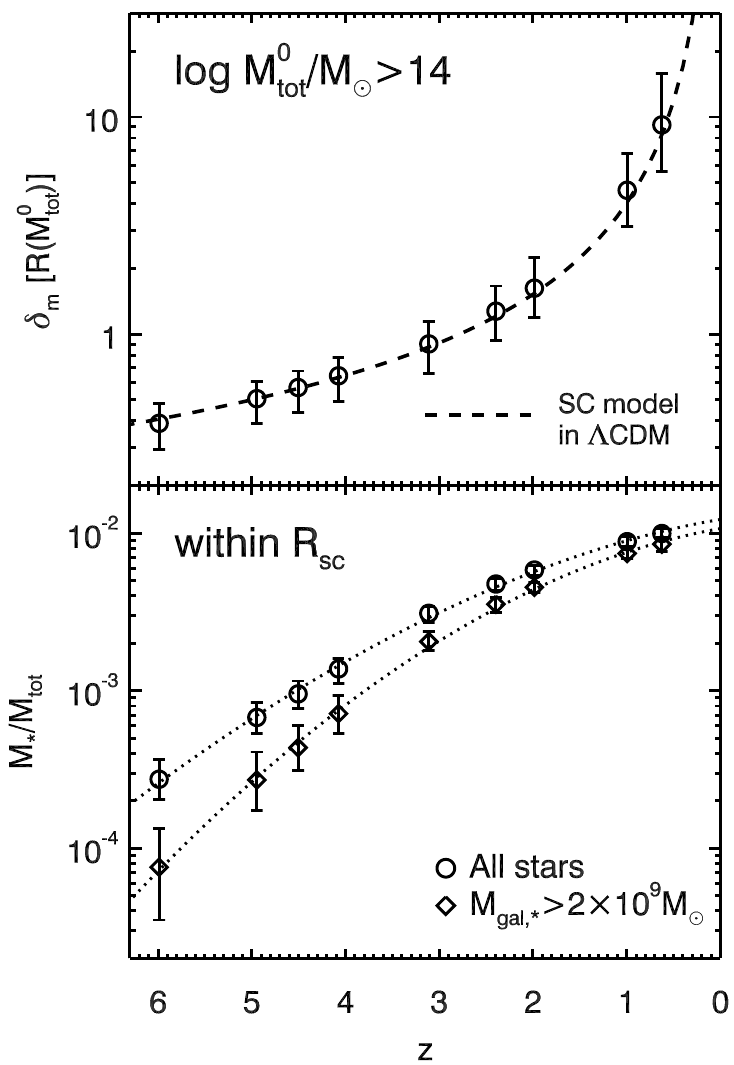}
\caption{\textit{ Top:} 
Matter overdensity inside the radius enclosing the final mass of protoclusters ($M^0_{\rm tot}>10^{14}\,\msun$) as a function of redshift. Dots are the medians, and scatter bars show $16^{\rm th}-84^{\rm th}$ percentile distributions. Dashed line is the critical matter overdensity $\delta_{\rm m}^{\rm sc}$ for collapse at $z=0$ predicted by the spherical top-hat collapse model in the $\Lambda$CDM universe. 
\textit{Bottom:} Ratio of stellar mass to total mass 
within the protocluster regions whose mean overdensity is equal to the critical value $\delta_{\rm m}^{\rm sc}$ of the $\Lambda$CDM cosmology. 
Open circles 
show the ratios computed from entire stellar mass, and open diamonds 
are calculated from the galaxies with $M_{{\rm gal}, \star}>2\times10^9\,\msun$. The dotted curves are the fitting functions given in Equation~\ref{eq:fit_mstar_mtot}.
}
\label{fig:f_mstar_delta}
\end{figure}

On the other hand, the SC model does not have an exact analytic solution in a flat universe with a non-zero cosmological constant, i.e., $\Omega_{\rm m}+\Omega_{\Lambda}=1$. We thus numerically solve the second order non-linear differential equation of the spherical top-hat overdensity $\delta_{\rm m}^{\rm sc}$ given in \citet{pace10}:
\begin{equation}
 \ddot{\delta}_{\rm m}^{\rm sc}+\left( \frac{3}{a}+\frac{\dot{E}(a)}{E(a)} \right)\dot{\delta}_{\rm m}^{\rm sc}-\frac{4}{3}\frac{(\dot{\delta}^{\rm sc}_{\rm m})^2}{1+\delta_{\rm m}^{\rm sc}}-\frac{3}{2}\frac{\Omega_{\rm m}}{a^5E^2(a)}\delta_{\rm m}^{\rm sc}(1+\delta_{\rm m}^{\rm sc})=0,
 \label{eq:delta_lcdm}
\end{equation}
where the derivatives are with respect to the expansion factor $a$, and $E(a)=H(a)/H_0=\sqrt{\Omega_{\rm m}/a^3+\Omega_{\Lambda}}$, where $H(a)$ and $H_0$ are the Hubble parameter at the epoch of an expansion factor $a$ and $z=0\,(a=1)$, respectively. 
The density parameters of $\Omega_{\rm m}=0.3$ and $\Omega_{\Lambda}=0.7$ are adopted in this calculation. 
We numerically search for the initial conditions $\delta_{\rm m}^{{\rm sc},i}$ and $\dot{\delta}_{\rm m}^{{\rm sc},i}=\delta_{\rm m}^{{\rm sc},i}/a_i$ at $a_i=10^{-3}$ that lead to $\delta_{\rm m}^{\rm sc}\rightarrow\infty$ at $z=0$, and find a solution $\delta^{{\rm sc},i}_{\rm m}=2.16\times10^{-3}$. The evolution of $\delta_{\rm m}^{\rm sc}$ is shown in Figure~\ref{fig:f_mstar_delta} (dashed line). For the general flat universe with non-zero $\Omega_{\Lambda}$, a fitting formula for the numerical solution of the SC model for the objects collapsing at $z=0$ is given in \ref{sec:EdS}.

\subsubsection{Overdensity of the \hr\ regions to be collapsed}

The SC model gives insight into the evolution of overdensities based on a simple assumption of homogeneous density distribution in a spherical region. However, in the real universe, structures are generally not spherical nor homogeneous. To examine if the simple assumption is applicable to practical cases, we compare the critical overdensity predicted by the SC model with the actual overdensity of the spherical region at a high redshift that encloses $M_{\rm tot}^0$, the total mass of each cluster at $z=0$ measured in the \hrl\ simulation.
The sphere is centered at the most massive galaxy among all the cluster progenitors at the redshift. 

The open circles in the upper panel of Figure~\ref{fig:f_mstar_delta} show the mean matter overdensity within the radius $R(M^0_{\rm tot})$ from the most massive progenitor of each of 189 \hrl\ clusters. 
It should be noted that $\delta_{\rm m}[R(M^0_{\rm tot})]$ for \hr\ clusters agrees quite well with the prediction of the SC model (dashed line) at all redshifts in the flat $\Lambda$CDM universe. This result demonstrates that the SC model is remarkably accurate in the $\Lambda$CDM universe at the mass scale of galaxy clusters, and thus the critical density threshold is applicable to identify protocluster regions.

\begin{figure*}
\centering 
\includegraphics[width=0.95\textwidth]{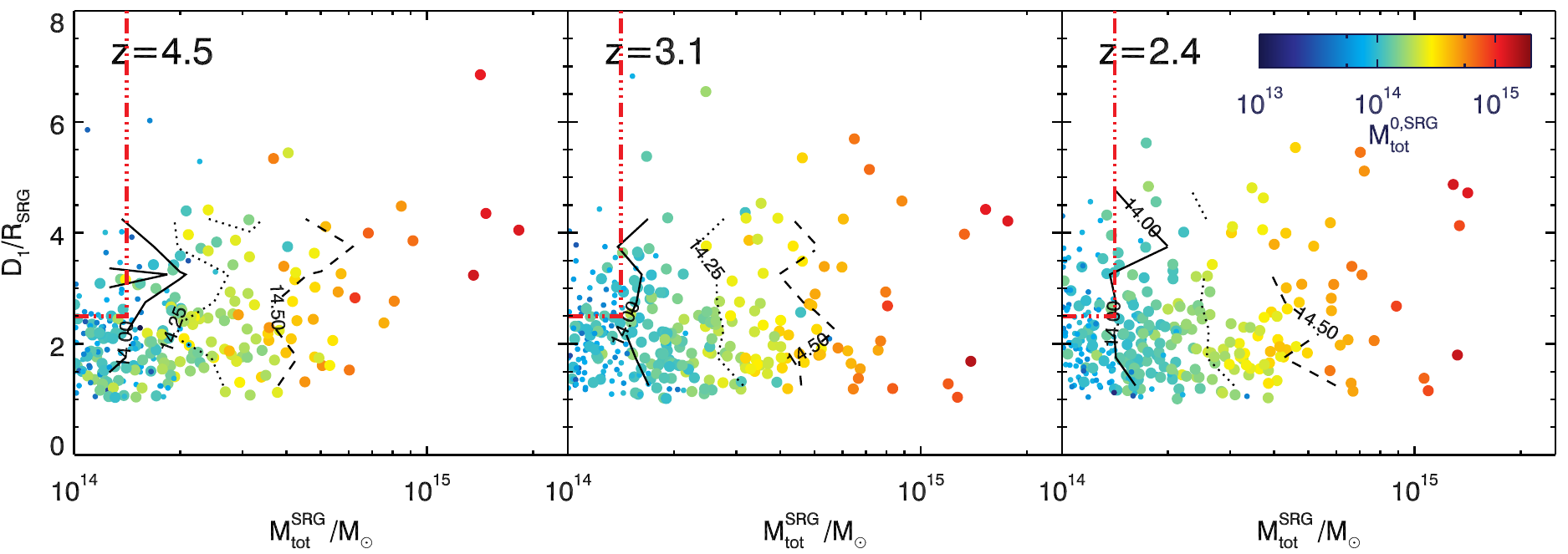}
\caption{Final total mass $M_{{\rm tot}}^{\rm 0,SRG}$ (encoded by color) as functions of distance to the nearest spherical region group (SRG) $D_1$ (normalized by $R_{\rm SRG}$) and its total mass $M^{\rm SRG}_{\rm tot}$ at each redshift. Each dot indicates a SRG, and color represents the final mass $M_{\rm tot}^{\rm 0,SRG}$ measured as in Equation~\ref{eq:PM}. 
Larger dots are the SRGs with cluster-scale mass ($M_{\rm tot}^{\rm 0,SRG}\ge10^{14}\,\msun$). 
The black solid, dotted, and dashed curves delineate the region of average final mass of $M_{\rm tot}^{\rm 0,SRG}=10^{14}$, $10^{14.25}$, and $10^{14.5}\,\msun$, respectively. 
The SRGs in the upper left corner demarcated by red lines are discarded in this work as protocluster candidates.
}
\label{fig:m_sc_f_dist}
\end{figure*}

\subsubsection{Identification of the regions enclosing protoclusters}

We have shown a good agreement between the spherical top-hat overdensity predicted by the SC model
and that actually measured for the \hr\ clusters.
However, 
to propose a protocluster identification scheme applicable to observations, it is necessary to find the relation between the total mass and stellar mass at the cluster mass scale. 
For the clusters with $\log M_{\rm tot}^0/\msun>14$ the bottom panel of Figure~\ref{fig:f_mstar_delta} shows the stellar-mass to total-mass ratio 
$M_\star/M_{\rm tot}$
within the spherical region having the critical overdensity of $\delta_m^{\rm sc}$ at redshift $z$. 
Open diamonds are the ratio when only the stars of the galaxies with $M_{{\rm gal}, \star} >2\times10^9\,\msun$ are used, and open circles are those when all stars are taken into consideration. We provide a fitting formula for the stellar-total mass relation in the following form:
\begin{equation}
\log M_\star^{\rm }/M_{\rm tot}^{\rm } =\alpha(1+z)^\beta+\gamma.
 \label{eq:fit_mstar_mtot}
\end{equation}
This formula can fit the ratio well as a function of redshift with $(\alpha,\beta,\gamma)=
(-0.055,1.903,-1.915)$ when the galaxies of $M_{{\rm gal}, \star}>2\times10^9\,\msun$ are used 
(shown as the dotted curve fitting the diamonds in the bottom panel of Figure~\ref{fig:f_mstar_delta}). When all stellar components are used (open circles), the best fit is made with the parameter set $(-0.057,1.755,-1.855)$. 
We note that the stellar-to-total mass relation is insensitive to mass in the case of the proto-objects of $M_{\rm tot}^0>10^{13}\,\msun$. This is because the region having the mean overdensity $\delta_{\rm m}^{\rm sc}$ is typically so large that the ratio converges to a value at a given redshift. 
The stellar-to-total mass ratio relation can be changed if the parameters of subgrid physics regulating star formation activities are changed. Therefore, the relation needs to be calibrated based on observations.

The protocluster identification starts with finding the candidate regions that enclose protoclusters. At a given epoch, we visit galaxies starting from the most massive ones, and inspect the spherical volume centered at the galaxy. The radius of the sphere is increased until the overdensity drops to the critical value $\delta_{\rm m}^{\rm sc}$ at that epoch. If the total mass contained within the sphere exceeds $10^{14}\,\msun$, the galaxy can be assumed as a candidate for the center of a protocluster. The fitting formula in Equation~\ref{eq:fit_mstar_mtot} is used to convert the observed stellar mass to the total mass.

The central candidate galaxies do not always locate at the density peak of each sphere. Thus, we compute the center of mass (CM) from all the galaxies with $M_{{\rm gal}, \star}>2\times10^9\,\msun$ located inside the spherical regions. To find the most representative center of galaxy distribution, we iterate the identification process until the CM converges to $|\vec{x}_{i-1}-\vec{x}_{i}|<\epsilon,$, where $\vec{x}_i$ is the CM at $i^{\rm th}$ iteration. In this study, we adopt $\epsilon=0.25\,$cMpc for efficient searching since a smaller $\epsilon$ does not notably affect the results. A sphere is selected as a region enclosing protoclusters when it finally has $M_{\rm tot}\ge10^{14}\,\msun$ after the iteration process.

In dense environment, the separations between the centers of the protocluster candidates can be very small.
We combine a protocluster candidate region $i$ with another one $j$ if $D_{ij}/R_i<1.0$ or $D_{ij}/R_j<1.0$, where $D_{ij}$ is the distance between the centers and $R_i$ and $R_j$ are the radii of the spheres within which the mean overdensity meets $\delta_{m}^{\rm sc}$.
In this case, we define the most massive sphere as the central one, and accordingly, $M_{\rm tot}^{\rm }$ of the central one is set as the estimated total mass of a spherical region group (SRG). 

\subsubsection{Reliability of the Protocluster Identification Scheme}
\label{sec:reliability}

We assume that the objects in the spherical regions of protoclusters identified based on the SC model eventually form cluster-scale objects by $z=0$. We evaluate the reliability of this approach by comparing the total mass of an SRG ($M_{\rm tot}^{\rm SRG}$)
at a redshift $z$ with the mass $M_{\rm tot}^{\rm 0,SRG}$ that ends up being inside clusters at $z=0$.
The latter is estimated using the final total mass weighted by the stellar mass of the cluster progenitor galaxies found within the SRG as follows:
\begin{equation}
M_{{\rm tot}}^{\rm 0,SRG}=\sum_{i=0}^{n} M(G\cap P_i)/ M(P_i) \times M_{{\rm tot},i}^0,
 \label{eq:PM}
\end{equation}
where $G$ is the set of the galaxies enclosed by an SRG, $P_i$ is the set of the progenitor galaxies of a cluster $i$, 
$M(P_{i})$ is the mass sum of $P_i$, 
and $M_{{\rm tot},i}^0$ is the final total mass of cluster $i$. 
The relation between $M_{{\rm tot}}^{\rm SRG}$ and $M_{{\rm tot}}^{\rm 0,SRG}$ tells us how reliably the spherical top-hat model predicts the final mass of enclosed objects.


It is reasonable to expect that the growth history of an SRG can be affected by its environment and the above relation may depend on the history. So we inspect if the final mass 
depends on
both $M_{{\rm tot}}^{\rm SRG}$ and mass growth environment. As a proxy of the environment, we choose $D_{1}/R_{\rm SRG}$, where $D_1$ is the distance to the nearest neighbor SRG and $R_{\rm SRG}$ is the radius of the target SRG. An SRG should have the total mass larger than half the total mass of the target SRG of interest to be qualified as a neighbor.

Figure~\ref{fig:m_sc_f_dist} shows the final mass $M_{{\rm tot}}^{\rm 0,SRG}$ (encoded by color of large circles) in the $D_{1}/R_{\rm SRG}$ versus $M_{{\rm tot}}^{\rm SRG}$ space. 
Redder color indicates larger final mass. Small dots are the SRGs with $M_{\rm tot}^{\rm SRG}\ge 10^{14}\,\msun$ at redshift $z$ but with $M_{\rm tot}^{\rm 0,SRG} \le 10^{14}\,\msun$ at $z=0$, namely failed protocluster candidates. 
The figure demonstrates a tight correlation of $M_{{\rm tot}}^{\rm SRG}$ with $M_{{\rm tot}}^{\rm 0,SRG}$, 
which justifies our use of the spherical overdensity criterion for identifying the protocluster centers. In particular, 90\% of the SRGs whose $M_{{\rm tot}}^{\rm SRG}$ is larger than $10^{14.2}\,\msun$ end up having $M_{{\rm tot}}^{\rm 0,SRG}> 10^{14}\,\msun$, indicating that they probably contain the authentic protoclusters. This illustrates the high reliability of our identification scheme. This figure also demonstrates that the final mass to be included in clusters is rather independent of the environment represented by the nearest neighbor SRG distance. We find, however, that the purity slightly improves if we discard the isolated small-mass SRGs with $M_{{\rm tot}}^{\rm SRG}<10^{14.15}\msun$ and $D_{1}/R_{\rm SRG}>2.5$ (the region enclosed by double dot-dashed lines). Based on these criteria, we examine the purity and completeness of our approach in identifying the bona-fide protoclusters in \ref{sec:sc_performance}. We find that the identification scheme recovers the authentic protoclusters with high reliability. We also show in \ref{sec:rsd} that the redshift-space distortion (RSD) does not significantly affect the performance of the protocluster identification scheme.

\begin{figure*}
\centering 
\includegraphics[width=0.9\textwidth]{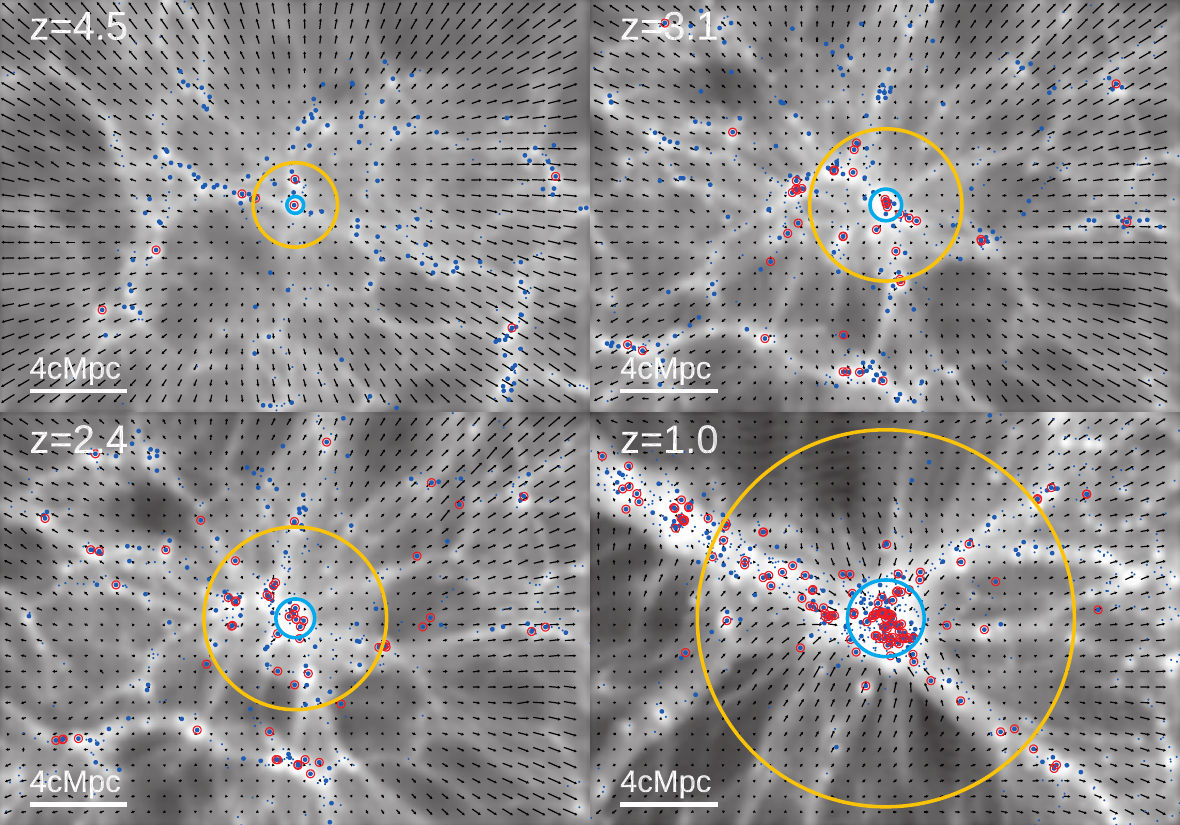}
\caption{Matter density and velocity fields within and in the vicinity of a protocluster at four epochs. Denser regions are brighter. The panels show matter distribution within $\pm2$cMpc from the most massive galaxy along the projected direction.  Blue and yellow circles indicate $R_{\rm vir}$ and \rturn\ measured from the density peak, respectively. All the blue dots are the gravitationally self-bound objects with the total mass greater than $10^{10}\,\msun$. Larger blue dots are the cluster progenitor objects, and among them, those with $M_{\star}>2\times10^9\,\msun$ are marked by red open circles.}
\label{fig:vfield}
\end{figure*}

\begin{figure}
\centering 
\includegraphics[width=0.45\textwidth]{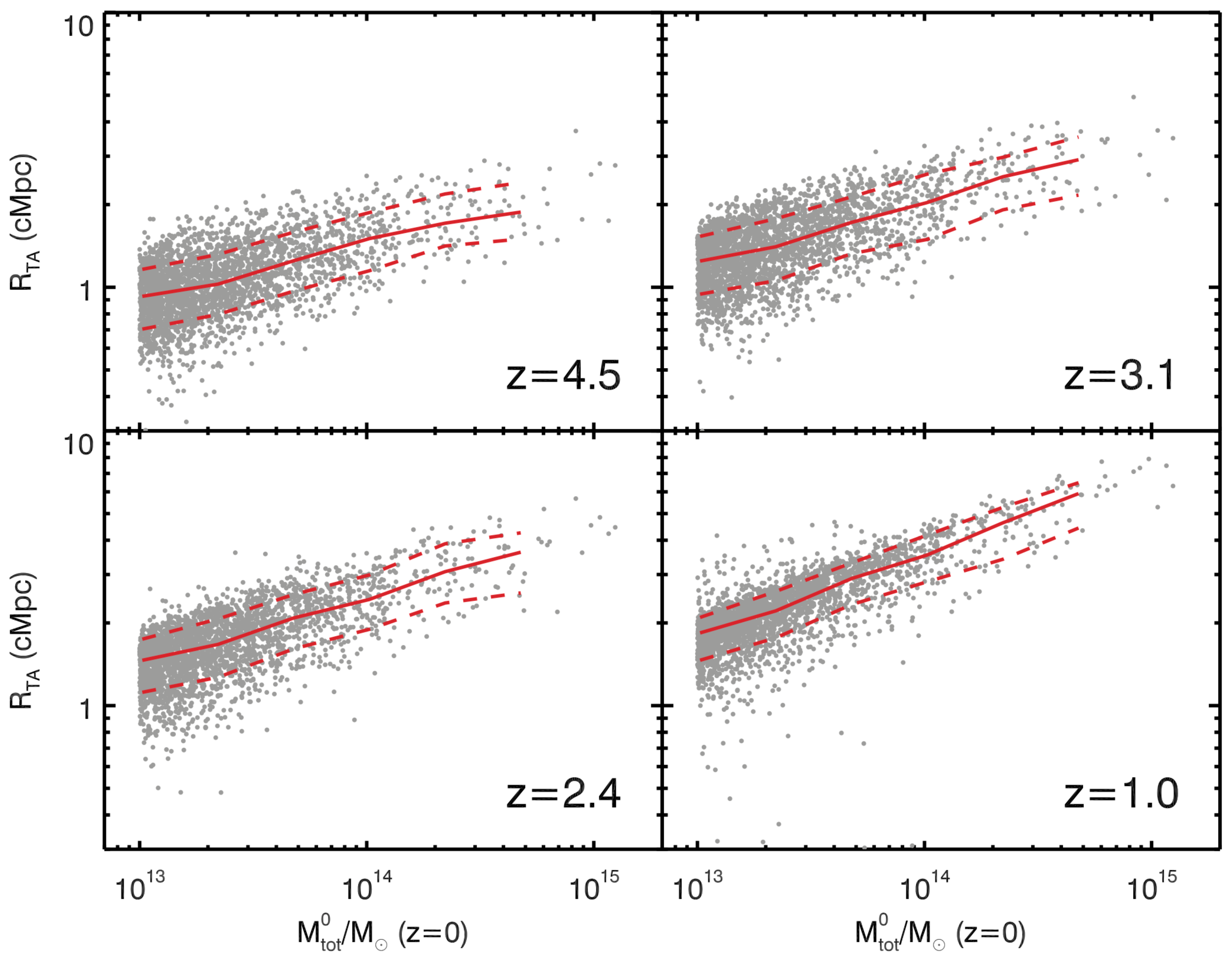}
\caption{Turnaround radius \rturn\ of the proto-objects as a function of the final total mass $M^0_{\rm tot}$ at $z=0$ that is measured in \hrl. Contrary to $R_{95}$, \rturn\ gradually increases as dense regions grow in mass and the Hubble parameter decreases with decreasing redshift.}
\label{fig:proto_rturn}
\end{figure}

\begin{figure*}
\centering 
\includegraphics[width=1\textwidth]{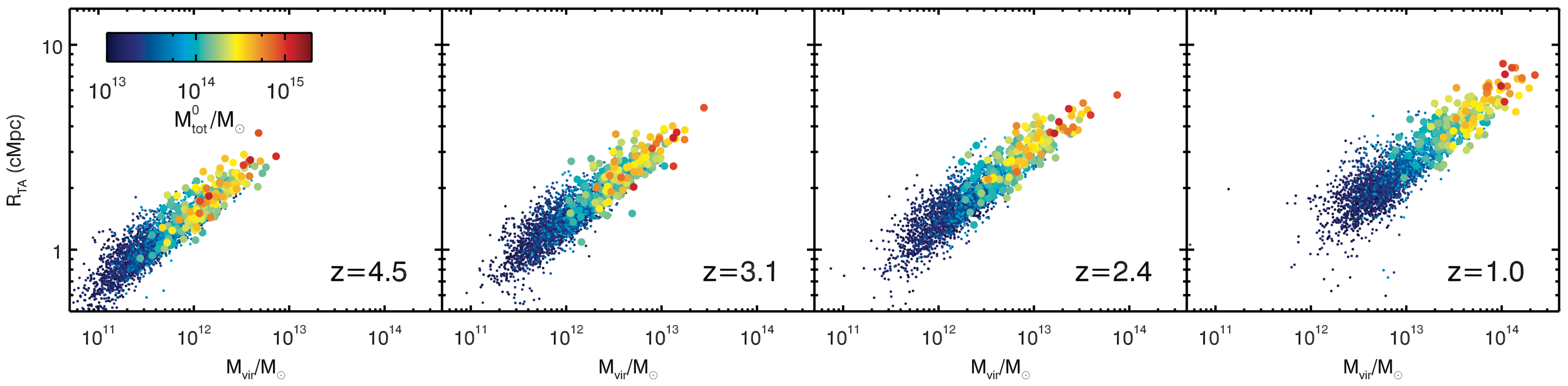}

\caption{Relations between the turnaround radius \rturn\ and virial mass $M_{\rm vir}$ at four epochs for the proto-objects that will have the final total mass of $M^0_{\rm tot}$ that is measured in \hrl. The final total mass is color-coded. 
Protoclusters are marked by large filled circles and non-protocluster objects are marked by small dots. 
}
\label{fig:m200_rturn}
\end{figure*}

\section{Protocluster member galaxies within Turnaround Radius}
\label{sec:turnaround_radius}

In numerical simulations and theories, it is relatively easy to define a protocluster as a group of objects that eventually contracts and forms a cluster. As described in Section~\ref{sec:merger_trees_proto}, the progenitors of cluster galaxies can be traced using their merger trees in numerical simulations, and the corresponding protoclusters can be identified. 

However, as shown in Figure~\ref{fig:proto_r95}, the progenitor galaxies of clusters are widespread up to $\sim 30$ cMpc at high redshifts and it is not reasonable to adopt all the progenitor galaxies as the physically-associated members of protoclusters. Most observations identify protoclusters by finding sufficiently overdense regions of galaxies~\citep[see][and references therein]{overzier16}. However, there has been no consensus on the value of the overdensity defining the membership of protocluster galaxies. 

Applying the virial radius in identifying protocluster galaxies is not so desirable as protoclusters are supposed to be the objects still under the process of formation and virized regions of protoclusters tend to vanish quickly as redshift increases.

We thus propose to define the protocluster member galaxies as those within the zero proper velocity surface from protocluster center.
The distance from a density peak to the zero-velocity surface is dubbed the turnaround radius \rturn. The turnaround radius is 
the distance to the spherical surface on which the gravitational infall counterbalances the Hubble expansion~\citep{gunn72}. 
The turnaround radius provides a theoretically motivated overdensity for defining the protocluster region, and also makes protoclusters physical objects where their member galaxies can have some degree of conformity. 
In this section we present a scheme for finding \rturn\ from observed galaxy distribution.

\subsection{Turnaround Radius}

To measure \rturn\ from the protocluster centers in \hr, we construct the matter (dark matter, gas, and stars) density and peculiar velocity fields on a uniform grid with pixel size of $\Delta x=0.128\,$cMpc. The proper radial velocity $v_r$ at $\boldsymbol{r}_1$ relative to a local density peak at $\boldsymbol{r}_0$ is given as follows:
\begin{equation}
 v_r=H(z)|\boldsymbol{r}|+\boldsymbol{e_r}\cdot\boldsymbol{v},
 \label{eq:vr}
\end{equation}
where $\boldsymbol{r}=\boldsymbol{r}_1-\boldsymbol{r}_0$, $H(z)$ is the Hubble parameter at redshift $z$, $\boldsymbol{e_r}$ is the unit vector of $\boldsymbol{r}$, and $\boldsymbol{v}$ is the peculiar velocity at $\boldsymbol{r}_1$ relative to the mean velocity of matter within $|\boldsymbol{r}|$. 
The turnaround radius is measured by finding the radius of a shell on which the average $v_{r}$ becomes zero. 

As an illustration, Figure~\ref{fig:vfield} shows the matter-density and velocity fields of a \hr\ protocluster region at four redshifts. The blue and yellow circles indicate $R_{\rm vir}$ and \rturn, respectively, centered at the most massive galaxy in the field at each epoch. Arrows are the proper velocity vectors projected onto a 4 cMpc-thick slice centered at the galaxy. 
The overdensity of the protocluster increases with time, and consequently, both \rturn and $R_{\rm vir}$ increase with time too. It can be seen that $R_{\rm vir}$ contains only the very center of the protocluster and becomes uninterestingly too small at high redshifts. On the other hand, \rturn\ is much larger than $R_{\rm vir}$, does separate the inner collapsing region from the outer expanding space, and embraces the high density region of intersecting filaments of galaxies. In this sense \rturn\ defines the outer boundary of the protocluster and the galaxies within \rturn\ can be called its `members'. 
Even though protocluster members are identified only within a spherical region, their distribution is quite anisotropic as the region encloses connecting filaments. 

Figure~\ref{fig:proto_rturn} shows \rturn\ of the \hr\ protoclusters and protogroups at four redshifts as a function of their final total mass at $z=0$. \rturn\ has a good correlation with the final mass. The tightness of the correlation increases toward low redshifts. The linear Pearson correlation coefficient is 0.634 at $z=4.5$ and this increases to 0.81 at $z=1.0$ in the $\log R_{\rm TA} - \log M^0_{\rm tot}/\msun$ plane. 
We have also checked if the turnaround radii measured from the most massive galaxies in SRGs are accurate compared to those of bonafide protoclusters, and find that 
more than 80\% of the SRGs have $R_{\rm TA}$ identical to that of the bonafide protoclusters (\ref{sec:rt_srg}).

\subsection{Correlations between Turnaround Radius, Virial Mass, and Viral Radius}

In this section we study the general nature of the turnaround radius by inspecting its relation with the virial mass and radius. The turnaround radius is known to be 3-4 times the virial radius of massive objects in the local universe~\citep{mamon04,wojtak05,rines06,cuesta08,falco13}. The virial mass of an object is defined as $M_{\rm vir}=4\pi r_{\rm vir}^3\Delta_c \rho_{\rm c}/3$, where $r_{\rm vir}$ is the virial radius within which the mean matter density is $\Delta_c$ times the critical density of the universe $\rho_{\rm c}=3H^2/8\pi G$, where $H$ is the Hubble parameter at $z$ and $\Delta_c$ is computed using the fitting formula derived by \citet{bryan98} for the cosmology with $\Omega_\Lambda>0$:
\begin{equation}
\Delta_c=18\pi^2+82x-39x^2,
 \label{eq:delta_vir}
\end{equation}
where $x=\Omega_{\rm m}(z)-1$.

Meanwhile, the total mean radial velocity at $r$ from the center of a bound object is the sum of the Hubble expansion velocity and mean infall peculiar velocity: $\left<{v}_r\right>=H(z)r+\left<{v}_{\rm infall}(r)\right>$, where $\left<{v}_{\rm infall}(r)\right>$ is the averaged radial velocity of matter in a spherical shell at radius $r$.

In the region where the Hubble flow starts to dominate and the
total mean radial velocity becomes positive, \citet{falco14} found a good approximation for the infall velocity profile 
as follows:
\begin{equation}
\left<{v}_{\rm infall}\right>\approx av_{\rm vir}\left({ r\over r_{\rm vir}}\right)^{-b}, 
 \label{eq:v_infall}
\end{equation}
where $v_{\rm vir} =\sqrt{GM_{\rm vir}/r_{\rm vir}}$ is the circular velocity at $r_{\rm vir}$,
and $a$ and $b$ are free fitting parameters. The best fit values found are $a=0.8\pm0.2$ and $b=0.42\pm0.16$ at $z=0$ in the N-body simulations of a $\Lambda$CDM universe with $\Omega_m =0.24$ and $h=0.73$~\citep{falco14}. 
Since  $\left<{v}_r\right>=0$ at the turnaround radius, the ratio of \rturn\ to $r_{\rm vir}$ can be reduced to $R_{\rm TA}/r_{\rm vir}=(a\sqrt{\Delta_c/2})^{1/(b+1)}$ by combining the  equations above with $r=R_{\rm TA}$ and $\left<{v}_r\right>=0$. Thus, the ratio $R_{\rm TA}/r_{\rm vir}$ is expected to be $\sim 4.3$ and in the range of 3.1 -- 6.2 at $z=0$.

\begin{figure}
\centering 
\includegraphics[width=0.47\textwidth]{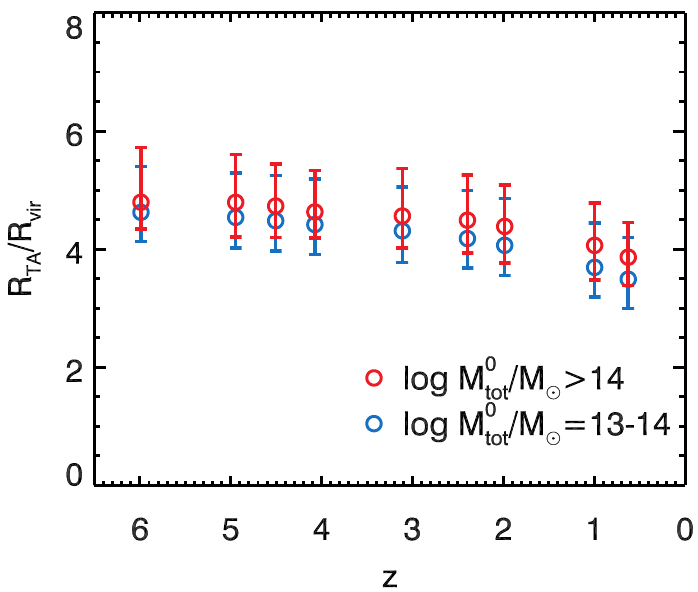}
\caption{Ratio of the turnaround radius \rturn\ to the virial radius  $R_{\rm vir}$ as a function of redshift. The blue and red circles correspond to the structures with $\log M^0_{\rm tot}/\msun=13-14$ and $\log M^0_{\rm tot}/\msun>14$ at $z=0$, measured from \hrl, respectively. The scatter bars show $16^{\rm th}-84^{\rm th}$ percentile distributions. This figure indicates that $R_{\rm TA}/R_{\rm vir}$ evolves very weakly before $z=2$.} 
\label{fig:ratio_rturn_r200}
\end{figure}

\begin{figure}
\centering 
\includegraphics[width=0.45\textwidth]{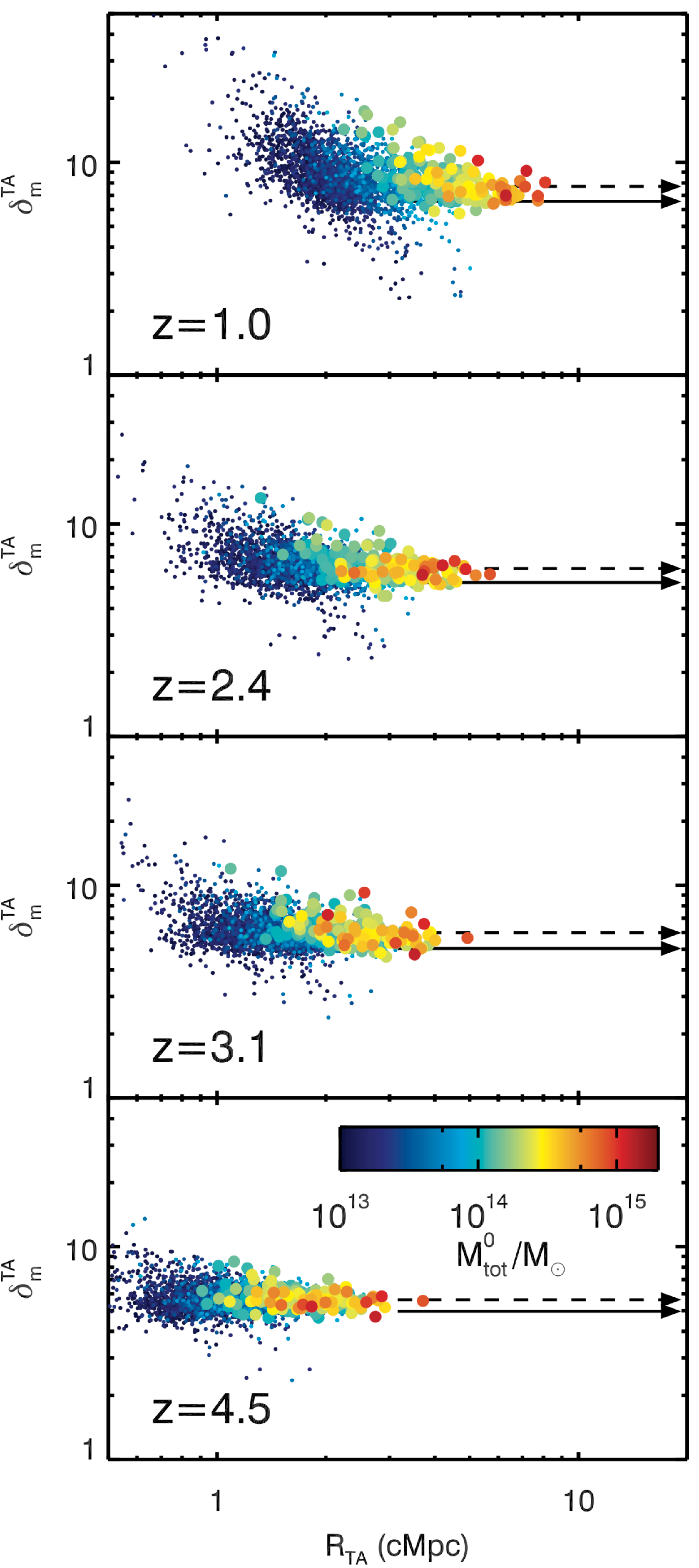}
\caption{Matter overdensity within the turnaround radius of proto-objects at the four redshifts. Protoclusters ($M^0_{\rm tot}\geq 10^{14}\,\msun$ at $z=0$, measured in \hrl) are marked by large filled circles and non-protocluster objects are marked by small dots. The color code presents the final total mass of proto-objects. The dashed and solid arrows indicate the medians and bottom 5\% of $\delta_{\rm m}^{\rm TA}$ of protoclusters. The matter overdensity of protoclusters only weakly increases from $\delta_{\rm m}^{\rm TA} \approx 5.0$ (bottom 5\%) at $z=4.5$ to $\delta_{\rm m}^{\rm TA}
\approx 5.3$ (bottom 5\%) at $z=2.4$ (see the text and Figure 12).}
\label{fig:delta_mass_rturn}
\end{figure}

\begin{figure*}
\centering 
\includegraphics[width=0.95\textwidth]{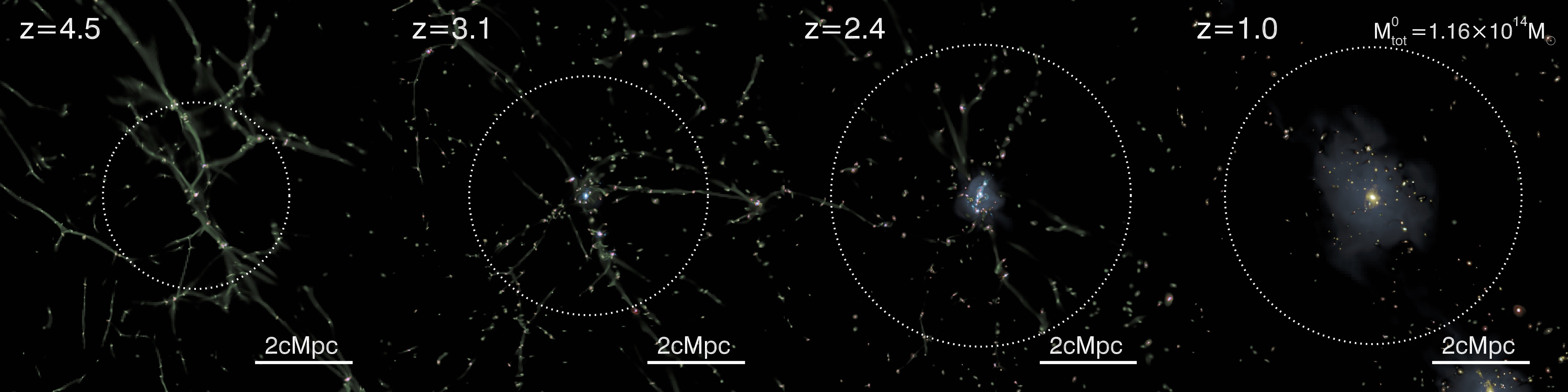}
\includegraphics[width=0.95\textwidth]{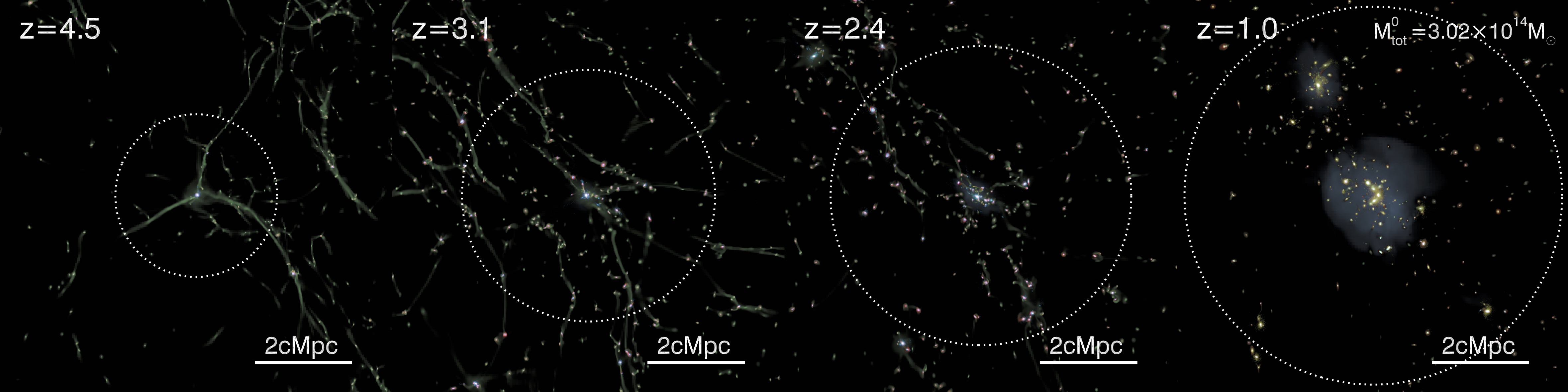}
\includegraphics[width=0.95\textwidth]{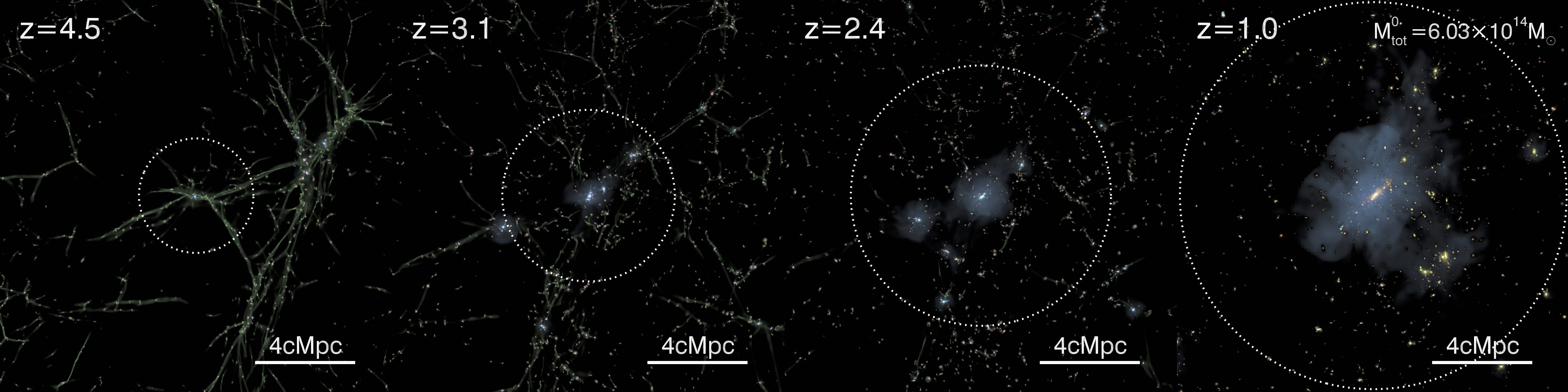}
\includegraphics[width=0.95\textwidth]{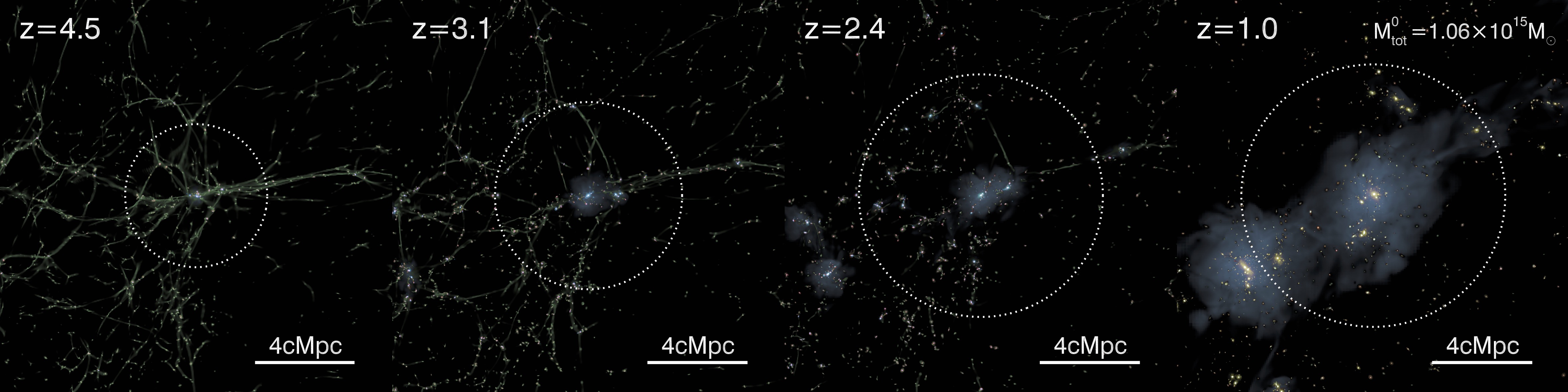}

\caption{Distribution of gas and stars in the regions of four protoclusters that end up forming clusters with $M_{\rm tot}^0\approx 10^{14}-10^{15}\,\msun$ at $z=0$ that is measured in \hrl. The dotted circles mark the turnaround radius of the protoclusters. Metal poor gas is colored in green, and gas color becomes redder with increasing metallicity. Younger stars are colored in blue and older ones are yellow. Grayish shades display the regions filled with the hot medium with $T>10^6$\,K. The upper two panels are relatively zoomed, as indicated by the scale bars.}
\label{fig:all_proj}
\end{figure*}

We now inspect the relation of \rturn\ with $M_{\rm vir}$ or $R_{\rm vir}$ 
directly for the \hr\ protocluster/group regions. Measurements are made relative to
the most massive galaxy in each region. 
Figure~\ref{fig:m200_rturn} demonstrates the tight correlation between \rturn\ and the virial mass at each epoch. Objects are distinguished in color according to their total mass at $z=0$. It can be noticed that the relation moves slowly downward with time and \rturn\ decreases at the same virial mass at lower redshifts. 

A weak evolution of the turnaround-to-virial radius ratio can be seen in Figure~\ref{fig:ratio_rturn_r200} for protoclusters (red, $M^0_{\rm tot}\ge10^{14}\,\msun$) and the proto-groups (blue, $M^0_{\rm tot}=10^{13}-10^{14}\,\msun$). The median of the ratio slowly decreases from 4.8 at $z=6$ to 3.9 at $z=0.625$ for protoclusters or clusters (red). The decreasing rate of the ratio is higher at $z<2$ than before as $\Delta_c$ significantly lowers. The ratio also decreases a little faster for proto-groups.
This seems to be caused by the disturbance of velocity field that becomes more severe for smaller mass objects at lower redshifts. The major origin of this weak redshift dependence will be discussed in the next section. 
Our measurement of $R_{\rm TA}/R_{\rm vir}$ at $z=0.625$ is consistent with the ratio range of 3.1-6.2 derived based on the semi-analytic approach of \citet{falco14}. 

\subsection{Matter Overdensity within Turnaround Radius}
\label{sec:delta_rturn}

The tight correlation between \rturn\ and $R_{\rm vir}$ implies nearly constant overdensity within \rturn\ at $z>2$. We measure the average matter overdensity of the \hr\ proto-objects inside the sphere of radius \rturn. Figure~\ref{fig:delta_mass_rturn} presents the matter overdensity $\delta_{\rm m}^{\rm TA}$ as a function of \rturn for all proto-objects. The large dots mark the protoclusters and small dots are proto-groups with the final mass of $M^0_{\rm tot}=10^{13}-10^{14}\,\msun$. 
The turnaround radius \rturn\ of protoclusters can temporarily decrease and $\delta_{\rm m}^{\rm TA}$ can jump up when they undergo close encounters with neighbors. In order to mitigate the impact of such temporal events, we choose to use the lower boundary (bottom 5\%) of the distribution of $\delta_{\rm m}^{\rm TA}$ shown in Figure~\ref{fig:delta_mass_rturn} for the threshold overdensity corresponding to \rturn. When protoclusters have close neighbors, the radius found with the lower boundary will be somewhat larger than the actual turnaround radius directly measured, and the protocluster regions are allowed to overlap. 
The bottom 5\% of the distribution of $\delta_{\rm m}^{\rm TA}$ are 4.96, 5.04, 5.30, and 6.55 at $z=4.5$, 3.1, 2.4, and 1.0, respectively. The median and $1\sigma$ dispersion are  
 5.63 ($\sigma=0.58$), 5.98 ($1.01$), 6.17 ($1.07$), and 7.71 ($1.91$), respectively. 

Like $R_{\rm TA}/R_{\rm vir}$, $\delta_{\rm m}^{\rm TA}$ also weakly evolves over time, with small scatter for protoclusters.
It should be noted that $\delta_{\rm m}^{\rm TA}$ hardly depends on \rturn\ or the final cluster mass of the protoclusters (large dots). 
On the other hand, $\delta^{\rm TA}_{\rm m}$ of the low mass structures with relatively small \rturn\ shows stronger evolution.
The scatter of $\delta_{\rm m}^{\rm TA}$ at small $R_{\rm TA}$ emerges when the field of interest is disturbed by neighbouring structures. 

Figure~\ref{fig:all_proj} illustrates the evolution of four \hr\ protoclusters representing different total mass scales at $z=0$. Dotted circles mark the turnaround radii, and properties shown are stellar mass density and age, gas density and metallicity. Similar to Figure~\ref{fig:vfield}, Figure~\ref{fig:all_proj} again shows that the volume within the turnaround radius does encompass the interesting large-scale structures connected to the protocluster cores.
It can be noticed in Figure~\ref{fig:all_proj} that $R_{\rm TA}$ is not always larger for the protoclusters with larger mass. It is also possible for $R_{\rm TA}$ to decrease temporarily when mergers happen. This is a desirable nature of $R_{\rm TA}$ as it is supposed to define the member galaxies of protoclusters and separate them from approaching nearby objects. However, during close interactions with neighbors, $R_{\rm TA}$ becomes smaller and $\delta^{\rm TA}_{\rm m}$ tends to increase. The upward scatter of the proto-objects in Figure~\ref{fig:delta_mass_rturn} can be attributed to such events.

\begin{figure}
\centering 
\includegraphics[width=0.47\textwidth]{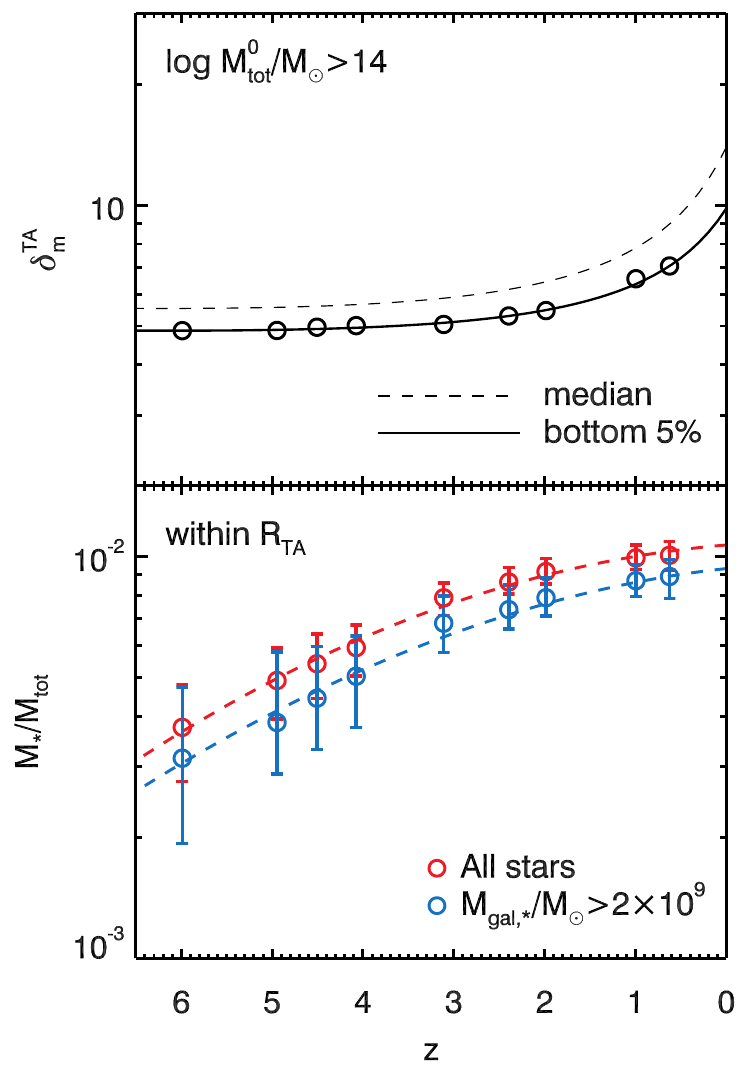}
\caption{{\it Top}: Redshift evolution of the overdensity within \rturn\ ($\delta_{\rm m}^{\rm TA}$) for the protoclusters of $M_{\rm tot}^0>10^{14}\,\msun$ that is measured in \hrl. The open circles indicate the bottom 5\% overdensity measured from the \hr\ protoclusters and the dashed and solid lines are the fits to the median and bottom 5\%. {\it Bottom}: ratio of stellar to total mass within \rturn\ as a function of redshift. The red and blue open circles denote the ratios measured from all stars and from galaxies with $M_{{\rm gal},\star}>2\times 10^9 \msun$, respectively. 
The dashed curves are the fits based on Equation~\ref{eq:fit_mstar_mtot} with fitting parameters given in section 4.4.} 
\label{fig:ratio_mstar_mtot_rturn}
\end{figure}

\subsection{Stellar mass to total mass conversion within Turnaround radius}
\label{sec:stellar_total_rturn}

We define the outer boundary of protoclusters as $R_{\rm TA}$, which is the turnaround radius enclosing the threshold overdensity given by Equation~\ref{eq:fit_delta_ta} below.
We will use an empirical relation between the total mass and stellar mass within $R_{\rm TA}$ so that the definition can be applied to observations.
Figure~\ref{fig:ratio_mstar_mtot_rturn} shows the redshift evolution of $\delta_{\rm m}^{\rm TA}$ of the \hr\ protoclusters (top) and the stellar-to-total mass ratio within the turnaround radius, 
$M^{\rm TA}_\star /M^{\rm TA}_{\rm tot}$, averaged over the \hr\ protoclusters. The stellar mass is obtained from all stars (red open circles) or only for the galaxies with $M_{{\rm gal},\star}>2\times10^9\,\msun$ (blue open circles).

The overdensity $\delta_{\rm m}^{\rm TA}$ delineating the bottom 5\% of the distribution at $z$ can be fit well by the following formula:
\begin{equation}
\delta_{\rm m}^{\rm TA}(z)=\frac{a\exp{(b\,(1+z)^{c})}}{(1+z)^{d}},
 \label{eq:fit_delta_ta}
\end{equation}
where $(a,b,c,d)=
(0.168,4.068,-0.381,-0.734)$,
which is shown as the solid line in the top panel of Figure~\ref{fig:ratio_mstar_mtot_rturn}. The error of the fit is smaller than 0.9\%.
As shown in Section~\ref{sec:delta_rturn}, $\delta_{\rm m}^{\rm TA}$ monotonically increases with time on average, and reaches a finite maximum at $z=0$. The evolution of $\delta_{\rm m}^{\rm TA}$ is weak at $z>2$, but becomes rapid at $z<1.5$ due to decrease of the Hubble parameter and disturbance by neighboring structures. 

\begin{figure}
\centering 
\includegraphics[width=0.45\textwidth]{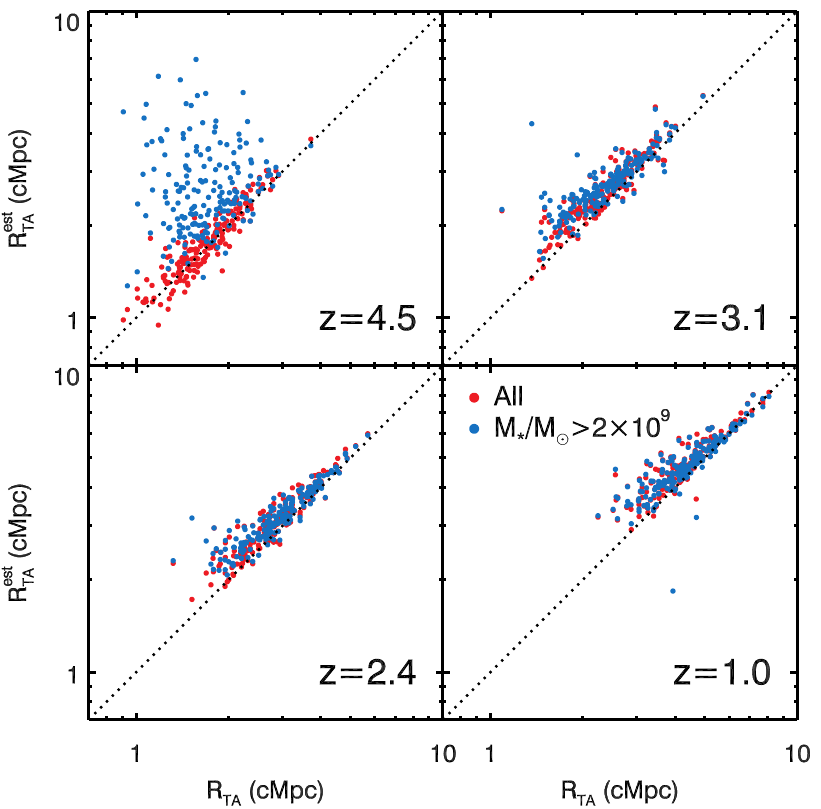}
\caption{The turnaround radius estimated from the stellar mass distribution using the relations shown in Figure~\ref{fig:ratio_mstar_mtot_rturn} versus the directly-measured turnaround radius of protoclusters. The former is based on the $\delta_{\rm m}^{\rm TA}$ of the bottom 5\%. The turnaround radii estimated by using all stars, and using only the galaxies more massive than $2\times10^9\,\msun$ are marked by red and blue dots, respectively. 
}
\label{fig:rturn_est}
\end{figure}

The stellar-to-total mass ratio within the turnaround radius can be also fit well by Equation~\ref{eq:fit_mstar_mtot} with $(\alpha,\beta,\gamma)=(-0.0092,2.027,-1.962)$ when all stellar mass is counted, or with $(-0.0128,1.882,-2.017)$ when only the stellar mass in the galaxies with $M_{{\rm gal},\star}>2\times10^9\,\msun$ are used. 
We use this fitting formula to derive the total mass from the stellar mass within a radius from each protocluster center, and find the radius within which the mean total mass density reaches the predicted $\delta_{\rm m}^{\rm TA}$ at the given redshift (i.e. Equation~\ref{eq:fit_delta_ta}). This gives the estimated turnaround radius.

Figure~\ref{fig:rturn_est} compares the directly measured $R_{\rm TA}$ with $R_{\rm TA}^{\rm est}$ estimated from stellar mass. They correlate quite well for both cases when all stellar mass is counted in $R_{\rm TA}^{\rm est}$ or only the stellar mass in the galaxies with $M_{{\rm gal},\star}>2\times10^9\,\msun$ is used. $R_{\rm TA}^{\rm est}$ tends to be larger than $R_{\rm TA}$ as expected, particularly for relatively smaller mass protoclusters, because we use bottom 5\% $\delta_{\rm m}^{\rm TA}$.
At $z=4.5$, when protoclusters have only a few galaxies above our stellar mass threshold, the correlation breaks. This necessitates to include the small mass galaxies with $M_{{\rm gal},\star} < 2\times10^9\,\msun$ at $z \gtrsim 4$ for accurate estimation of $R_{\rm TA}$ and reliable identification of protocluster environment.

\section{Summary and Discussion}
\label{sec:summary}

In this paper we have proposed a practical method to find galactic protoclusters in observational data, and demonstrated its validity to the protoclusters in the cosmological hydrodynamical simulation \hr. We first define `protoclusters' as galaxy groups whose total mass within $R_{\rm vir}$ is currently less than $10^{14}\msun$ at their epochs  but would exceed that limit by $z=0$.  Conversely, `clusters' are the groups of galaxies whose virial mass currently exceeds $10^{14}\,\msun$. Therefore, there can be a mixture of clusters and protoclusters at $z>0$. The extent of a protocluster is defined as the spherical volume within the turnaround radius or the zero-velocity surface. The future mass that a protocluster would achieve at $z=0$ is estimated using the spherical top-hat collapse model. The whole concept is schematically visualized in Figure~\ref{fig:definition}. 

Our protocluster identification method is summarized as follows:

1. Visit galaxies starting from the most massive ones, and measure the mean total mass density within radius $R$. The total mass is obtained from the stellar mass by using the conversion relation in Equation~\ref{eq:fit_mstar_mtot}.

2. Find the radius where the mean density drops to the threshold density given by the SC model. Equation~\ref{eq:fit_delta_lcdm} is a useful fitting formula for the threshold overdensity $\delta_{\rm m}^{\rm sc}$.

3. Adopt the galaxy (or nearby density peak) as a protocluster center candidate if the total mass included within the radius is greater than $10^{14}\,\msun$. Group the spherical regions if their separation is less than their radii. Protocluster centers are now identified.

4. The protocluster region is defined as the spherical volume from the protocluster center up to the turnaround radius. The turnaround radius is the radius where the mean overdensity drops to the threshold value given by Equation~\ref{eq:fit_delta_ta}. The stellar mass to total mass conversion within $R_{\rm TA}$ is made using Equation~\ref{eq:fit_mstar_mtot}, with the parameters given in Section~\ref{sec:stellar_total_rturn}.

\


\begin{figure}
\centering 
\includegraphics[width=0.45\textwidth]{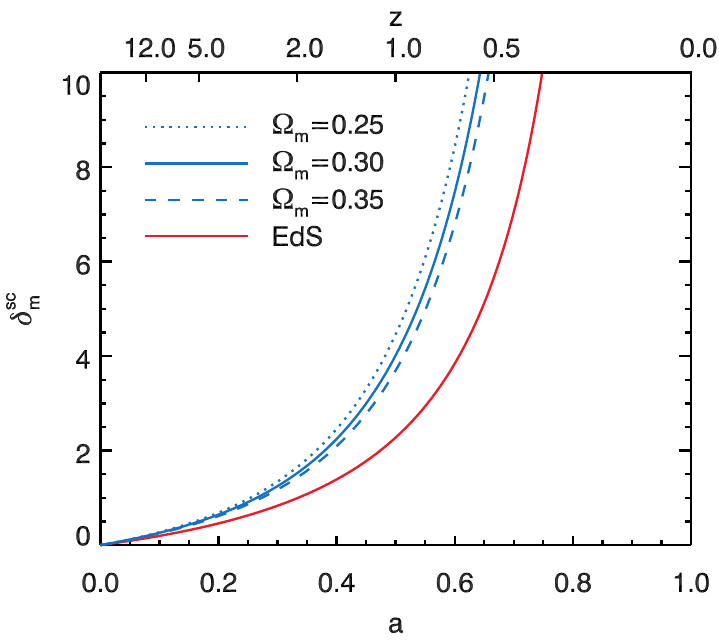}
\caption{The critical overdensity for complete collapse at $z=0$ 
given by the spherical top-hat collapse model in the Einstein-de Sitter  universe (EdS, red) and the flat $\Lambda$CDM universes with three different matter density parameters.}
\label{fig:delta_EdS_LCDM_multi}
\end{figure}

\hr\ used in this paper adopts a flat $\Lambda$CDM cosmology with $\Omega_{\rm m}=0.3$ and $\Omega_{\Lambda}=0.7$.
As the threshold density given by the spherical top-hat collapse model is used to find the protocluster centers, it will be useful to check
how sensitive the threshold is to the cosmology adopted.
We examine how $\delta_m^{\rm sc}$ changes depending on the matter density parameter while keeping the geometry of the universe flat and fixing the dark energy equation of state parameter to $-1$. Our choice of \hr\ is based on the Planck data~\citep{planck16}. This is close to the recent measurement of \citet{dong23} who used the extended Alcock-Paczyński test to obtain 
$\Omega_m=0.285_{-0.009}^{+0.014}$. In Figure~\ref{fig:delta_EdS_LCDM_multi}, $\delta_{\rm m}^{\rm sc}$ for four choices of $\Omega_{\rm m}$, i.e., $0.25, 0.3, 0.35$, and 1, safely bracketing the recent observational values, are plotted.  
The figure shows that $\delta_{\rm m}^{\rm sc}$ differs on average only by $\sim$12\% at $z=6-2$ among the flat $\Lambda$CDM models with $\Omega_{\rm m}$ from 0.25 to 0.35. Therefore, the threshold density used for finding the protocluster centers is not very sensitive to the choice of the matter density parameter, when the current tight constraint on the parameter is taken into account. 

To estimate the reliability of this prescription, we use the clusters at $z=0$ with $M^0_{\rm tot}\ge 10^{14}\,\msun$ and groups with $10^{13}\,\msun < M^0_{\rm tot} < 10^{14}\,\msun$  identified in \hrl, a low-resolution version of \hr. There are 2,794 objects with $M^0_{\rm tot}>10^{13}\,\msun$ in the zoomed region of \hr, and among them 189 are clusters. Merger trees are constructed for these objects, and all progenitor galaxies are identified. We apply our protocluster identification scheme to the galaxy distributions at four simulation snapshots of $z=4.5, 3.1, 2.4$, and 1, being motivated by the ODIN survey of Lyman-$\alpha$ emitters.
We find a tight correlation between the mass within the protocluster regions identified in accordance with the SC model, and the final mass to be situated within clusters at $z=0$. In particular, it is highly likely (probability $\gtrsim$ 90\%) for a protocluster region to evolve to a cluster if the region contains a total mass greater than about $2\times10^{14}\msun$, meaning that the region is likely to be the authentic protocluster.

We have defined the outer boundary of protoclusters as the zero-velocity surface at the turnaround radius. 
Even though protocluster members are identified within a spherical region, their distribution is quite anisotropic as the region encloses numerous filaments beaded with galaxies. 
The definition would make sense if the galaxies 
within the turnaround radius do share some physical properties, which is not found for those outside. In the next study, we will examine the physical properties and evolution of the protocluster galaxies based on the definition proposed in this study.

\section*{acknowledgments}
J.L. is supported by the National Research Foundation of Korea (NRF-2021R1C1C2011626). C.P. and J.K. are supported by KIAS Individual Grants (PG016903, KG039603) at Korea Institute for Advanced Study. BKG acknowledges the support of STFC through the University of Hull Consolidated Grant ST/R000840/1, access to {\sc viper}, the University of Hull High Performance Computing Facility, and the European Union’s Horizon 2020 research and innovation programme (ChETEC-INFRA -- Project no. 101008324).
This work benefited from the outstanding support provided by the KISTI National Supercomputing Center and its Nurion Supercomputer through the Grand Challenge Program (KSC-2018-CHA-0003, KSC-2019-CHA-0002). This research was also partially supported by the ANR-19-CE31-0017 \href{http://www.secular-evolution.org}{http://www.secular-evolution.org}. This work was supported by the National Research Foundation of Korea(NRF) grant funded by the Korea government (MSIT, 2022M3K3A1093827).
Large data transfer was supported by KREONET, which is managed and operated by KISTI. 
This work is also supported by the Center for Advanced Computation at Korea Institute for Advanced Study. 

\appendix
\label{sec:appendix}

\section{Structure Finding and Merger Trees}
\label{sec:merger_trees}

We use a galaxy finder \pgalf\ introduced by \cite{kim+22} to extract self-bound and stable galaxies from the snapshots of \hr. \pgalf\ is devised to identify the Friend-of-Friend group of particles from the distribution of heterogeneous particles, i.e., star, MBH, gas, and dark matter in \hr. For the mixture of various types of particles, \pgalf\ uses an adaptive linking length to connect a pair of particles of different species or masses. \pgalf\ identifies self-bound substructures in the FoF halos. We classify a substructure as a galaxy when it contains stellar particles. To find galaxies from a FoF halo, \pgalf\ first constructs an adaptive stellar density field and hierarchically determine the membership of the particles bound to the galaxies centered at stellar density peaks. A bound particle is eventually assigned to a galaxy when it is located inside the tidal boundary of the galaxy. We note that a galaxy identified in this process is generally composed of heterogeneous particles. For the substructures with no stellar particles, a similar process is conducted for the rest matter species. For a full description on the method, refer to \cite{kim+22}.

Since stellar or dark matter particles carry their own unique identification numbers (IDs) throughout the simulation runs, we are able to trace the progenitors/descendants of substructures between two time steps. A branch of a merger tree is described using the binary relation between the two sets of all stellar particles in two snapshots, motivated by the Set theory. First, we define $\mathcal{S}_i$ as a set of all stellar particles at time step, $t_i$. Then, 
\begin{equation}
    \mathcal{S}_i = \mathcal{S}_{i-1} \cup \{s| {\textrm{new stars born in\;}} (t_{i-1}, t_i]\},
\end{equation}
where ``new stars'' are those created between time steps, $t_{i-1}$ and $t_i$. We define $\mathcal{G}_i^j$ as the group of star particles of the $j$'th galaxy at time step $i$. Because a stellar particle is never destroyed in \hr, $\mathcal{S}_{i-1}\subseteq \mathcal{S}_i$. Our galaxy finder dictates that $\mathcal{G}_i^j \cap \mathcal{G}_i^k = \varnothing$ for $j\neq k$. The relation below is also satisfied;  $\bigcup\limits_{j=1}^{n}\mathcal{G}_i^j\subseteq \mathcal{S}_i$, where $n$ is the total number of galaxies identified in time step $i$. The left-hand and right-hand sides of the equation are not always equal due to unbound stray stellar particles which are not bound to any galaxies.

We associate galaxies between two snapshots by mapping a set of stellar particles (a galaxy) at a time step into sets of stellar particles (galaxies) at the next time step using \ysamtm~\citep{jung14,leej14}. In \ysamtm, we define the $j$'th galaxy as the main descendant of the $k$'th galaxy when satisfying the mapping:
 \begin{equation}
f: k \underset{\textrm{desc}}{\mapsto}  
\underset{j}{\arg \max}[P(\mathcal{G}_{i+1}^j|\mathcal{G}_{i}^k)],
 \end{equation} 
where $P(\mathcal{G}_{i+1}^j|\mathcal{G}_{i}^k)$ is the fractional number of stellar particles of the $k$'th galaxy to be found in the $j$'th galaxy. Multiple galaxies in time step $i$ are allowed to have a common main descendant in time step $i+1$ once the mapping is satisfied or in short $f(j) = f(k)$ for $j\ne k$.

Now we consider the reverse mapping as
\begin{equation}
g: k \underset{\textrm{prog}}{\mapsto}  
\underset{j}{\arg \max}[P(\mathcal{G}_{i-1}^j|\mathcal{G}_{i}^k)],
\end{equation}
which denotes that the $j$'th galaxy in time step $i-1$ is the main progenitor to $k$'th galaxy in time step $i$. Unlike the mapping $f$ for the main descendant, in principle, multiple galaxies in time step $i$ cannot have a common main progenitor in time step $i-1$. So, in this case $g(j) \ne g(k)$ for all $j\ne k $. This is because we assume that a galaxy cannot be fragmented into multiple descendants in \ysamtm.

The mapping $f$ is the {\it left} inverse mapping of $g$ ;  it can be defined more formally as,
\begin{eqnarray} \label{fg}
(f\circ g)(j) &\equiv& f(g(j)) = j, \\ \label{gf}
(g\circ f)(j) &\equiv& g(f(j)) \ne j.
\end{eqnarray}
Here, equation (\ref{fg}) means that the main descendant of a main progenitor is the galaxy itself. One the other hand, $g$ is not {\it left} inverse mapping of $f$ (Eq. \ref{gf})  because of the case when the $j$'th galaxy is merged into its descendant.

Our tree building scheme does not allow two galaxies to have the same main progenitor (or $g(j) \neq g(k)$ for $j\neq k$), but this usually happens when a galaxy flies by a more massive galaxy. To circumvent such cases, we remove the main progenitor mapping of the less massive galaxy (the flying-by one) and trace back its previous history until its actual main progenitor is found, using the most bound particle (MBP). The MBP is a particle that has the largest negative total energy in the galaxy~\citep{hong16} and, thus, we assume that the MBPs trace density peaks of galaxies. We use dark matter particles as the MBPs because, unlike stellar particles, they do not disappear when backtracking snapshots. We also use the MBP scheme to trace the substructures with no stellar particles. The merger trees of substructures are constructed by connecting the progenitor-descendant relations across the all snapshots. The progenitor/descendant relation of FoF halos is traced based on the merger trees of their most massive substructures. Further details of the tree buliding algorithm are given in~\citet{park22}.

\section{Identification of Cluster Progenitors in HR5}
\label{sec:cluster_candidates}

\begin{figure}
\centering 
\includegraphics[width=0.45\textwidth]{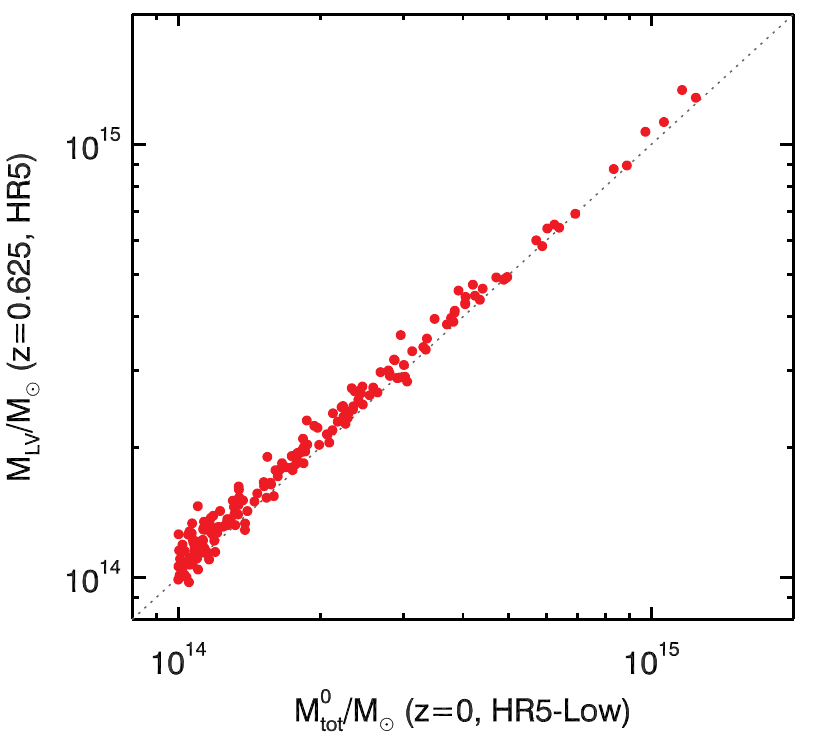}
\caption{The relation between the total mass of the clusters found at $z=0$ in \hrl\ and the LV mass at $z=0.625$ in \hr. The LV mass is on average $\sim6$\% higher than the cluster mass due to the matter that are contained in voxels at the epoch but will not form the clusters.}
\label{fig:mass_relation_tot}
\end{figure}

\begin{figure}
\centering 
\includegraphics[width=0.45\textwidth]{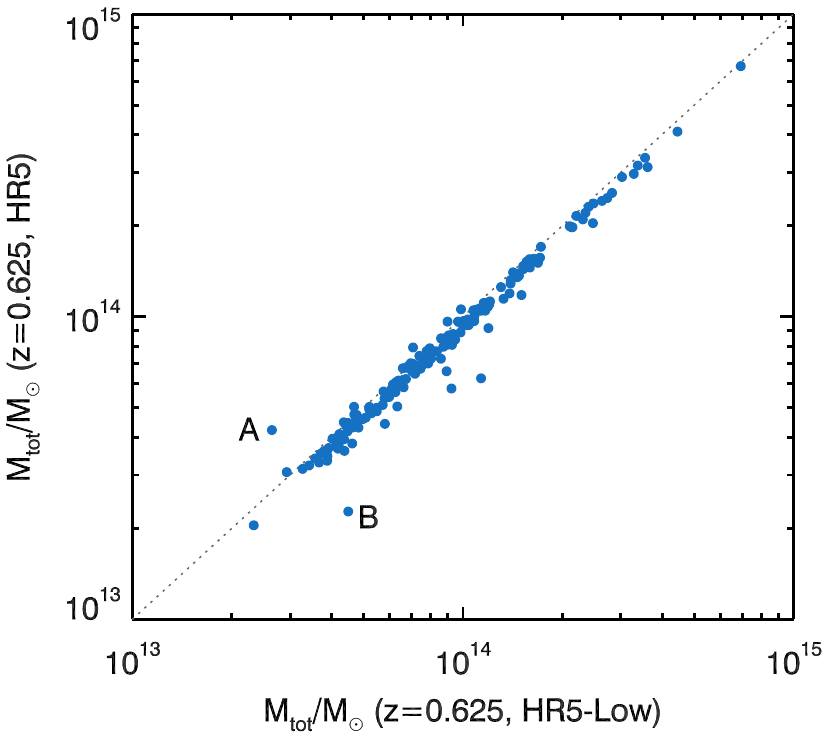}
\caption{Relation between the total mass of the main progenitors ($z=0.625$) of the clusters found at $z=0$ in \hrl\ and the total mass of their counterparts in \hr. Halo \texttt{A} is the one that is identified as two separate structures in \hrl\ while a smaller one already becomes a substructure of the halo in \hr. Halo \texttt{B} is the opposite case. The halos in \hr\ are $\sim9\%$ less massive than their counterparts in \hrl\ because their small neighboring structures are not well resolved in \hrl. 
}
\label{fig:mass_relation}
\end{figure}

In this section, we describe the details of the identification process of the clusters in \hr\ using its low resolution simulation \hrl. 
While \hr\ achieves a spatial resolution down to $\Delta x\sim1\,$kpc and minimum dark matter particle mass of $m_p \simeq 6.89\times10^7\,\msun$, \hrl\ is set to have a spatial resolution down to $\Delta x \sim16\,$kpc with a minimum dark matter particle mass of $m_p \simeq 3.02\times10^9\,\msun$. Because the main purpose of \hrl\ is to identify structures at $z=0$, we use the parameters and initial conditions of \hr\ without any modification or calibration. We identify structures from the snapshot at $z=0$ and 0.625 of \hrl\ using \pgalf. At $z=0$, we find 2,794 halos in $M^0_{\rm tot}\ge10^{13}\,\msun$ and 189 halos in $M^0_{\rm tot}\ge10^{14}\,\msun$ with the number fraction of lower level particles less than 0.1\%, which ensures the mass contamination lower than 0.7\%. The dark matter particles of the clusters are traced back to $z=0.625$ using their IDs, to search for the progenitors of halos of $M^0_{\rm tot}\ge10^{13}\,\msun$.

We measure the LV of a cluster in terms of the Cartesian grids. In \hrl, we place a mesh of uniform cubic grids with $\Delta l=0.512\,$cMpc over the entire volume of interest (the simulated zoomed region). To build a density field, we use the dark matter particles of cluster halos at $z=0$. When dark matter particles in a grid do not belong to (or are not members to) a single cluster, the grid is finally associated with the cluster which contributes most to the grid mass. By utilizing the LV method with the \hrl\ data, we are able to define protocluster regions at an arbitrary redshift.

\begin{figure}
\centering 
\includegraphics[width=0.45\textwidth]{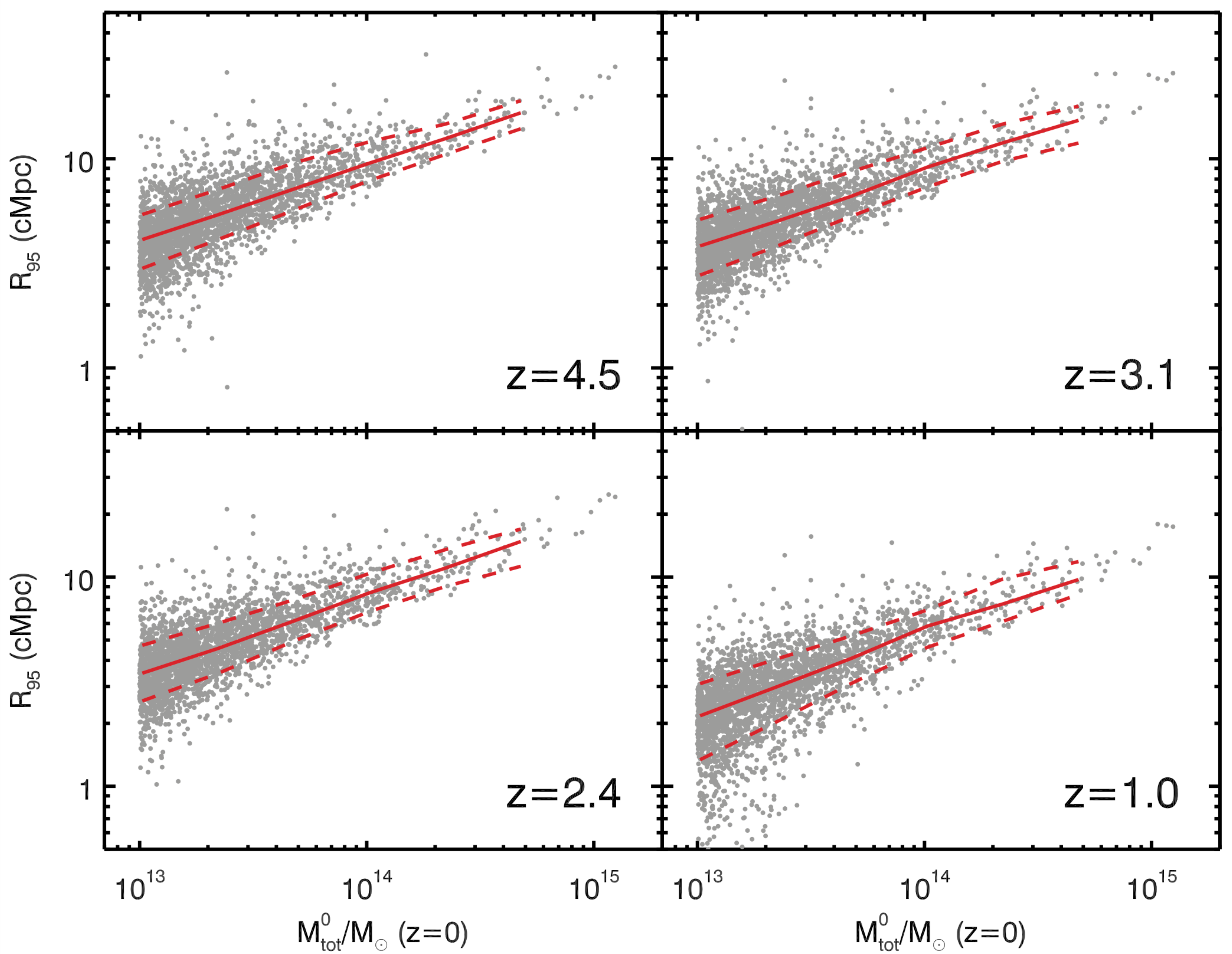}
\caption{Radius that encloses 95\% of the stellar mass of the proto-objects of the FoF halos identified at $z=0$ as a function of their final mass that is measured from \hrl. The radius measurement is centered at the most massive galaxy in each proto-object. Red dashed and sold lines mark $16^{\rm th}$ and $84^{\rm th}$ percentiles and the median of $R_{95}$ at a given final mass. 
}
\label{fig:proto_r95}
\end{figure}

In the subsequent analysis we assume that the LVs of \hr\ clusters are identical to the LVs of corresponding \hrl\ clusters. In the last snapshot of \hr, therefore, we are able to find structures inside the LVs directly imported from the \hrl\ clusters. 
We only use grids having mass larger than $10^{10}\,\msun$ because 97.5\% of galaxies with $M_\star\ge10^9\,\msun$ have $M_{\rm tot}>10^{10}\,\msun$. This mass cut helps us minimize the contamination by non-cluster progenitors in the LVs of the cluster progenitors at $z=0.625$.

Figure~\ref{fig:mass_relation_tot} presents the relation between the cluster mass in \hrl\ at $z=0$ and the corresponding LV mass $M_{\rm LV}$ in \hr\ at $z=0.625$. The two masses are nearly same with a median scattering of $\sim6\%$. The mass difference may be caused by matter that happens to be enclosed in the LVs but would not fall into the cluster at $z=0$.

To examine the consistency or similarity in particle distributions between \hrl\ and \hr\ especially on halo scales at $z=0.625$, we identify an \hr\ FoF halo which is spatially closest to the main progenitor of each \hrl\ cluster. Here, the progenitor of a cluster is determined by the scheme described in Section 2.2.

Figure~\ref{fig:mass_relation} shows the relation of FoF halo masses between the main progenitors of clusters in \hrl\ and their counterparts in \hr\ at $z=0.625$. Except for two cases marked by \texttt{A} and \texttt{B}, all FoF halos in the two simulations have nearly same mass. We slightly overestimate the mass of FoF halos in \hrl\ compared to \hr\ because of the purer mass resolution which tends to more easily destroy clumpy  structures in the outskirts of halos. Here, \texttt{A} and \texttt{B} are the cases when substructures are distinguishable only within either\hr\ or \hrl. We assume that, although rare, the adaptive linking length may cause the different FoF halo identification between two simulations at different resolutions. Alternatively, the different-resolution simulations may, of course, produce different particle distributions more often in the outskirts of halos especially around a close binary or a multiple system of halos.

Figure~\ref{fig:proto_r95} shows $R_{95}$, the radius enclosing 95\% of stellar mass in cluster progenitors, as a function of the final total mass. The progenitors of more massive halos tend to have larger $R_{95}$. The range of $R_{95}$ is consistent with \citet{muldrew15} who measure $R_{90}$ of protoclusters using a semi-analytic model of \citet{guo11}. In this study, we suggest the turnaround radius as the physical size of protoclusters, instead of $R_{\rm 95}$ because $R_{\rm 95}$ measures merely the spatial extent of the distribution of progenitor galaxies.

\section{Spherical top-hat overdensity in the $\Lambda$CDM and Einstein de-Sitter universe}
\label{sec:EdS}

\begin{figure}
\centering 
\includegraphics[width=0.45\textwidth]{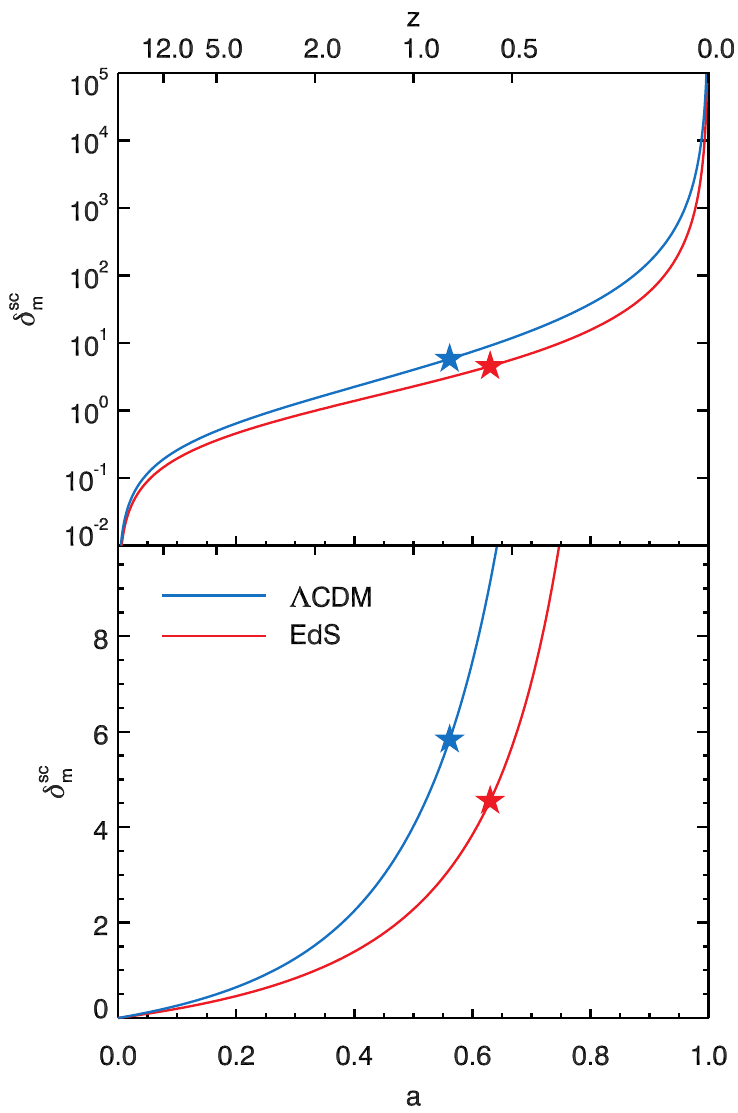}
\caption{The critical overdensity of a homogeneous top-hat sphere collapsing at $z=0$ predicted by the spherical top-hat collapse model in the $\Lambda$CDM (blue) and EdS (red) universe in logarithmic (top) and linear (bottom) scales. Stars indicate the epoch and overdensity when the sphere reaches its maximum radius in each universe.}
\label{fig:delta_EdS_LCDM}
\end{figure}

In the Einstein de-Sitter (EdS) universe with $\Omega_{\rm m}=1$, the outermost radius $R$ of a sphere of mass $M$ evolves over time $t$ as follows:
\begin{equation}
 \ddot{R}=-\frac{GM}{R^2},
 \label{eq:spt_EdS}
\end{equation}
where $G$ is the gravitational constant. This equation has the cycloidal solution:
\begin{equation}
 t=\frac{t_{\rm max}}{\pi} (\theta-\sin\theta), \\
 R=\frac{R_{\rm max}}{2} (1-\cos \theta),
 \label{eq:sol_spt}
\end{equation}
 where $t_{\rm max}$ is the time when the sphere reaches a maximum radius $R_{\rm max}$. In this solution, the spherical region collapses at the collapse time $t_{\rm c}=2t_{\rm max}$ ($\theta=2\pi$). The overdensity of the sphere at a given epoch derived from the analytic solution is given by~\citep[e.g.,][]{peebles80,suto16}:
\begin{equation}
 \delta_{\rm m}^{\rm sc}\equiv\frac{\rho_{\rm m}^{\rm sc}}{\bar{\rho}_{\rm m}}-1=\frac{9}{2}\frac{(\theta-\sin\theta)^2}{(1-\cos\theta)^3}-1.
 \label{eq:delta_EdS}
\end{equation}
A homogeneous density sphere that collapses at $z=0$ reaches its maximum radius at $z=0.59$ with $\delta_{\rm m}^{\rm sc}=9\pi^2 /16 -1\simeq 4.55$ in the EdS universe. 
For comparison, the linear theory predicts overdensity $\delta_m^{\rm lin}\simeq1.062$ at $t_{\rm max}$ in the EdS universe.

In the flat universe with non-zero $\Omega_{\Lambda}$, the expansion factor of maximum radius $a_{\rm max}$ can be derived using the formula~\citep{peebles84,eke96}:
\begin{equation}
 a_{\rm max}=\left[ \frac{\exp{(3\sqrt{\omega}I(\omega))}-1}{2\sqrt{\omega}}\right]^{\frac{2}{3}}\exp{(-\sqrt{\omega}I(\omega))},
 \label{eq:aturn_lcdm}
\end{equation}
where $\omega=\Omega_{\Lambda}/\Omega_{\rm m}$ and $I(\omega)$ is given from:
\begin{equation}
 I(\omega)=\frac{1}{2}\int^{a_{\rm c}}_{0}\frac{\sqrt{a}}{\sqrt{\omega a^3+1}} {\rm d}a,
 \label{eq:Iomega}
\end{equation}
where $a_{\rm c}$ is the expansion factor at the time of collapse. These equations give $a_{\rm max}=0.56$ in the case of $a_{\rm c}=1.0\,(z=0)$ and the overdensity at the epoch is interpolated as $\delta_{\rm m}^{\rm sc}=5.85$ for our choice of the $\Lambda$CDM universe. When $\Omega_{\rm m}=1.0$ and $\Omega_{\rm \Lambda}=0$, Equation~\ref{eq:delta_lcdm} has the solution that is equal to the exact solution of the EdS universe case derived above. 

Figure~\ref{fig:delta_EdS_LCDM} shows the overdensity evolution in a homogeneous sphere that collapses at $z=0$ in the EdS (blue) and $\Lambda$CDM (red) universe. The two filled stars indicate the overdensities at the epochs of maximum radius. Because dark energy counteracts gravitational collapse and the growth of overdensity is relatively slower, the sphere should have higher overdensity in the universe with $\Omega_{\Lambda}>0$ than in the EdS universe, to be able to collapse by $z=0$.

Since $\delta_{\rm m}^{\rm sc}$ does not have an exact analytic solution in the $\Lambda$CDM universe adopted, we find a formula that fits the numerical solution of the SC model 
for the objects collapsing at $z=0$:
\begin{equation}
\delta_{\rm m}^{\rm sc}(z)=\frac{0.0224\exp{(5.39\,z^{-0.246})}}{(1+z)^{0.294}}.
 \label{eq:fit_delta_lcdm}
\end{equation}
This formula has the error $<0.3$\% in the redshift range of $z=[0.5,6.0]$.

\section{Performance of the Protocluster Identification Scheme based on the SC model}
\label{sec:sc_performance}

\begin{figure*}
\centering 
\includegraphics[width=0.95\textwidth]{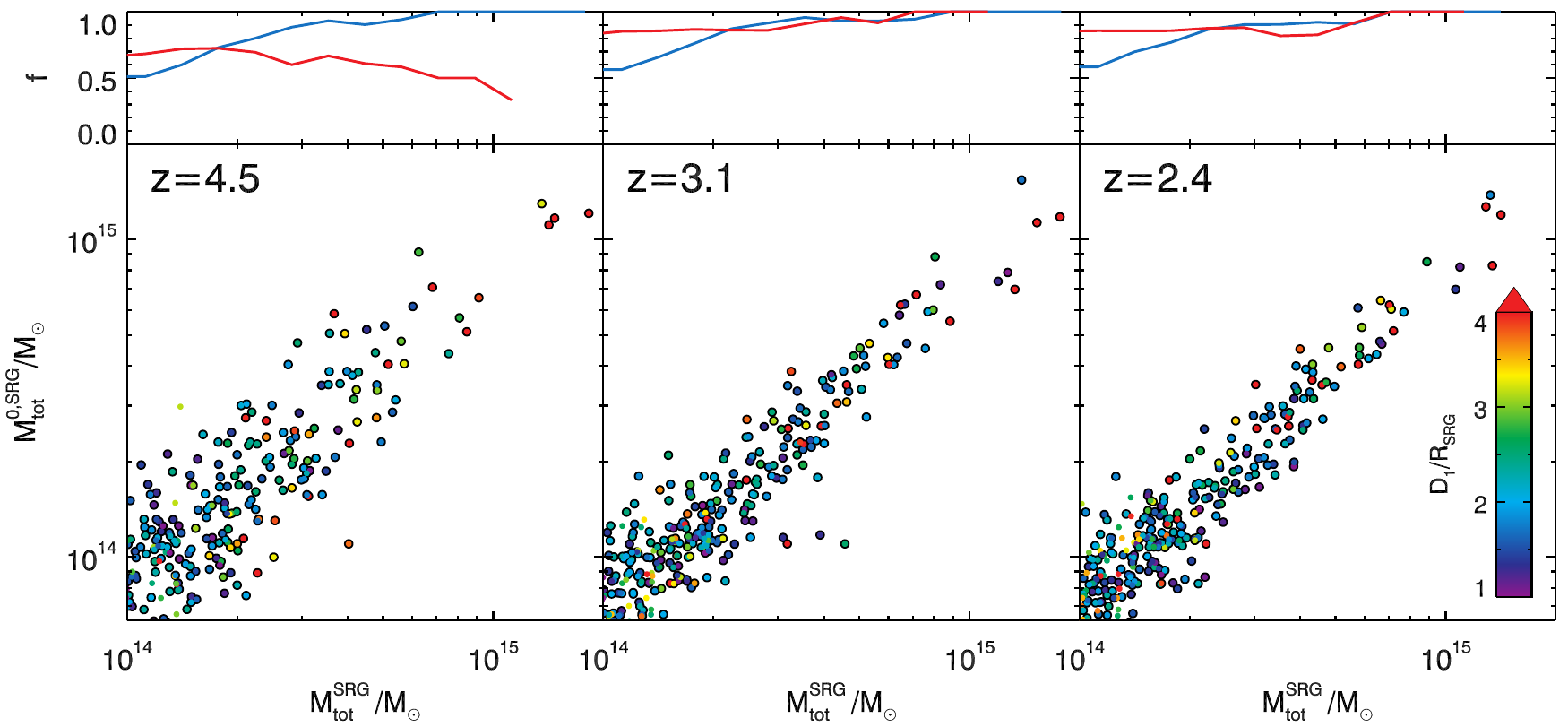}
\caption{{\it Bottom}: Final mass of SRGs estimated from the final mass of bona-fide protoclusters (those identified based on merger trees) embedded in the SRGs ($M_{\rm tot}^{\rm 0,SRG}$) as a function of the total mass of SRGs ($M_{\rm tot}^{\rm SRG}$). 
Color code denotes $D_1/R_{\rm SRG}$. Colored dots mark all the SRGs sample and black concentric circles indicate the SRGs with $M_{\rm tot}^{\rm SRG}>10^{14.15}$ or $D_1/R_{\rm SRG}<2.5$. As also seen in Figure~\ref{fig:m_sc_f_dist}, most protoclusters have $D_1/R_{\rm SRG}\lesssim4$.
We note that $M_{\rm tot}^{\rm 0,SRG}$ is an estimated mass to examine the prediction accuracy of $M_{\rm tot}^{\rm SRG}$. {\it Top}: Purity (blue) and completeness (red) of the bona-fide protoclusters in the spherical regions found by the SC model as a function of $M_{\rm tot}^{\rm SRG}$. The purity is the number fraction of the SRGs enclosing bona-fide protoclusters to the entire SRGs above a given mass. The completeness is the number fraction of the authentic protoclusters which are recovered by SRGs and more massive than a given mass.}
\label{fig:m_sc_m_final}
\end{figure*}

Figure~\ref{fig:m_sc_m_final} demonstrates the completeness and purity of our protocluster identification scheme (top) and the relation between $M_{\rm tot}^{\rm 0,SRG}$ and $M_{\rm tot}^{\rm SRG}$ (bottom). We define the purity as the number fraction of the SRGs enclosing bona-fide protoclusters (those identified based on merger trees) to the all SRGs more massive than a given mass. The completeness is the number fraction of the authentic protoclusters enclosed by SRGs above a given mass. In these statistics, we assume that an SRG recovers a protocluster when the most massive galaxy of the SRG is the member of the protocluster and half the galaxy mass of the protocluster is enclosed by the SRG. In this scheme, an SRG can be associated with only one protocluster.  
Color code in the bottom panels indicates the $D_1/R_{\rm SRG}$ parameter. Colored dots show the distribution of all the SRGs sample and black concentric circles mark the SRGs with $M_{\rm tot}^{\rm SRG}>10^{14.15}$ or $D_1/R_{\rm SRG}<2.5$.
These two different mass definitions are overall in good agreement, particularly at $z<4$. Their correlation becomes tighter with decreasing redshift as structures form and develop further. The completeness and purity show that more than $80\%$ of protoclusters can be recovered by our scheme with $\sim60\%$ purity at $z\sim2-3$. The purity increases to $80\%$ in $M_{\rm tot}^{\rm SRG}\ge2\times10^{14}\,\msun$. At $z=4.5$, however, these statistics are inevitably poorer than at lower $z$, because galaxies have not had time to develop yet. We note that the purity and completeness are enhanced by $\sim10\%$ if an SRG is allowed to associate with all the protoclusters in which half their galaxy mass is enclosed by the SRG.

\section{Redshift-Space Distortion Effect on the Protocluster Identification}
\label{sec:rsd}

\begin{figure*}
\centering 
\includegraphics[width=0.95\textwidth]{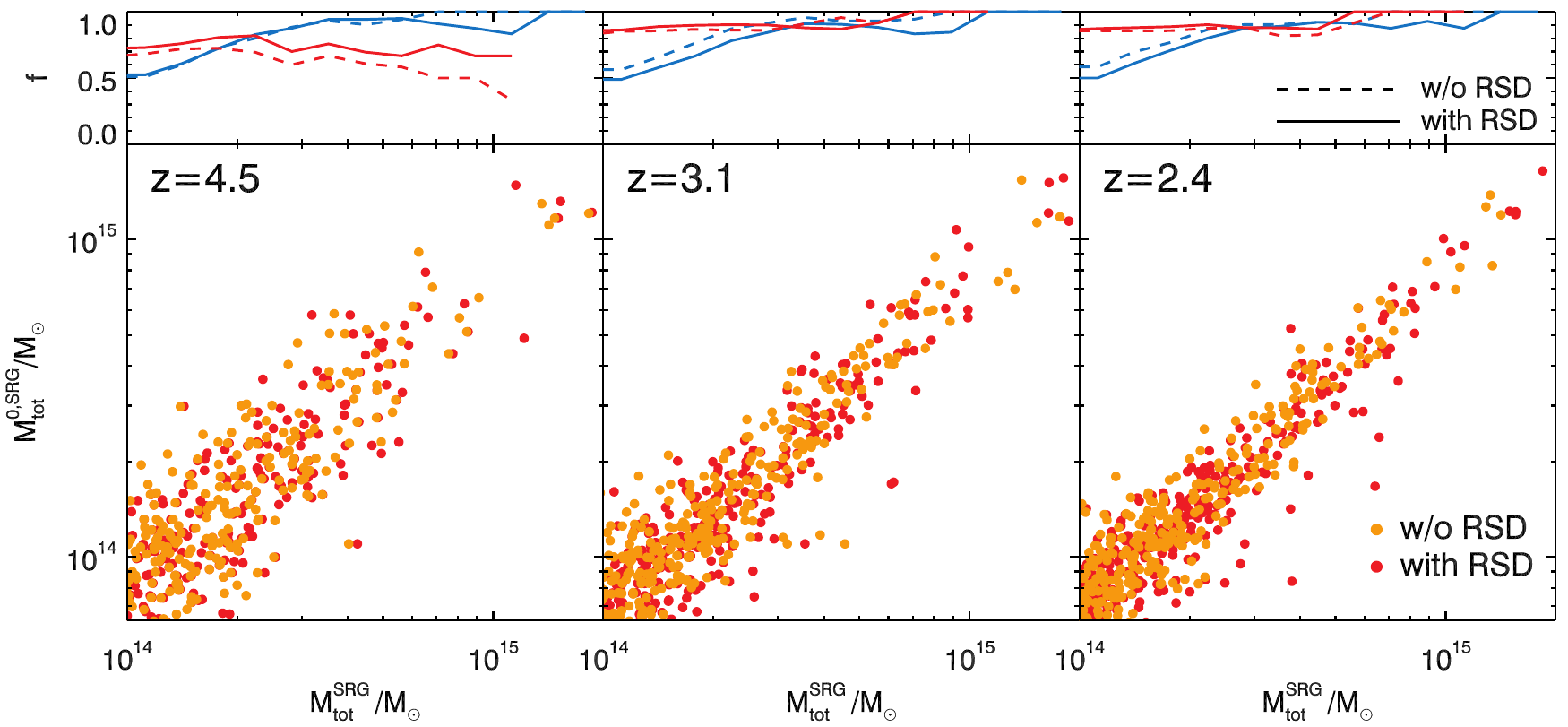}
\caption{Same as Figure~\ref{fig:m_sc_m_final}, but for the cases with and without the RSD effect. In the bottom panels, the scatter between $M_{\rm tot}^{\rm 0,SRG}$ and $M_{\rm tot}^{\rm SRG}$ is similar between the two cases with and without the RSD effect. Upper panels show that the RSD effect lowers the purity while it slightly enhances the completeness at given mass. This is caused by the RSD effect that makes overdense regions look flattened in the redshift space~\citep{kaiser87}, resulting in the overestimation of the SRG radius.}
\label{fig:m_sc_m_final_rsd}
\end{figure*}

The peculiar velocities of galaxies distort the distribution of galaxies in redshift space~\citep[e.g., see][]{guzzo97,hamilton98}. We examine the impact of RSD on the protocluster identification scheme. In this test, we assume that a virtual observer has the line of sight aligned with the major axis of the \hr\ zoom-in region. The redshift of a snapshot is assigned to the center of the zoomed region, and the cosmological redshifts of the galaxies in the snapshot are computed from the distance relative to a virtual observer at $z=0$. The Doppler redshifts induced by the peculiar velocities of galaxies are added to the cosmological redshifts, and the distances to the galaxies are re-estimated from the combined redshifts. The standard deviations of the differences between the intrinsic and redshift-distorted distances are 2.3, 3.0, and 3.6\,cMpc at $z=2.4$, 3.1, and 4.5, respectively. Figure~\ref{fig:m_sc_m_final_rsd} presents the impact of the RSDs on the protocluster identification scheme. Since the large-scale peculiar velocity vector tends to point toward overdense regions, the galaxy distribution near a protocluster is statistically flattened along the line of sight in redshift space~\citep{kaiser87}. This results in slight overestimation of the overdensity and size of the top-hat spheres of  dense regions. The final impact is that the completeness increases, at higher redshifts in particular, while the purity slightly decreases. The bottom panels of Figure~\ref{fig:m_sc_m_final_rsd} show that the RSD effect slightly increases the SRG mass, but overall distribution is similar between the cases with and without the RSD effects. These statistics are computed based on the assumption that an SRG is only associated with a protocluster. The purity and completeness can change if an SRG is allowed to recover multiple protoclusters. This result demonstrates that the RSD effect does not have significant impact on the protocluster identification scheme.

\section{Turnaround Radius of Spherical Region Groups}
\label{sec:rt_srg}

\begin{figure}
\centering 
\includegraphics[width=0.45\textwidth]{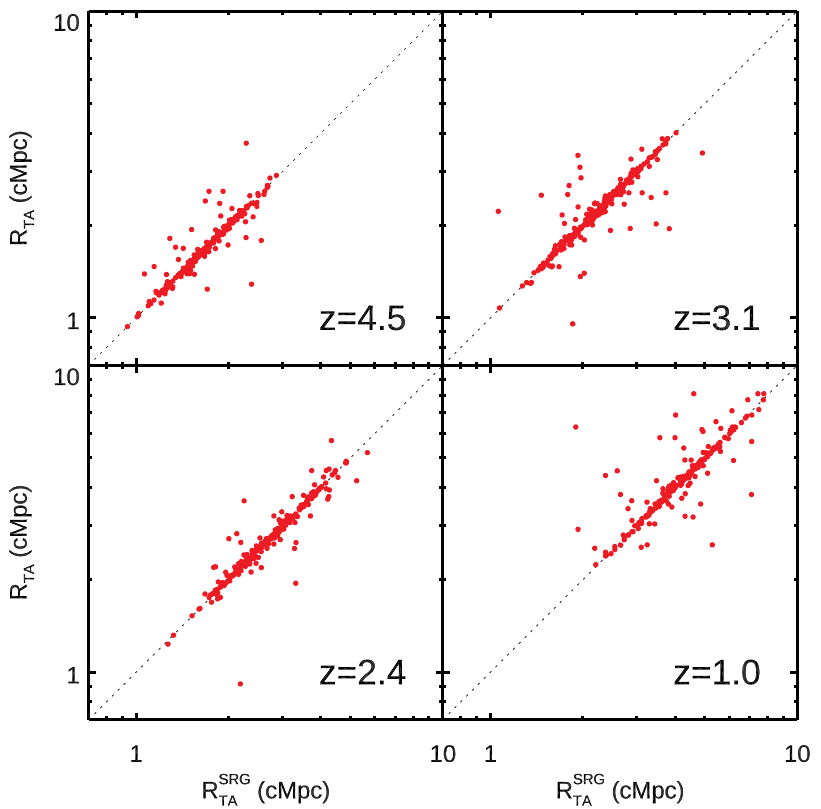}
\caption{Relation of turnaround radius of bona-fide protoclusters (\rturn) to turnaround radius measured from the most massive galaxies in SRGs ($R_{\rm TA}^{\rm SRG}$). A protocluster is assumed to be associated with an SRG when half its galaxy mass is enclosed by the SRG. Scatter is caused when the most massive galaxy of an SRG is not the most massive one of its host protocluster. We find that $\sim80\%$ SRGs recover the \rturn\ of enclosed protoclusters.   
}
\label{fig:rturn_srg}
\end{figure}

We examine if the turnaround radius is reasonably recovered in the SRGs. For consistency with the \rturn\ measurement for the bona-fide protoclusters, we measure the turnaround radius relative to the most massive galaxy in an SRG ($R_{\rm TA}^{\rm SRG}$), and compare it with \rturn\ of the protocluster that hosts the most massive galaxy of the SRG and share half its total galaxy mass with the SRG. Figure~\ref{fig:rturn_srg} shows \rturn$-R_{\rm TA}^{\rm SRG}$ relation at the four redshifts. In this comparison, more than 80\% of the SRGs have $R_{\rm TA}^{\rm SRG}$ identical to \rturn. The scatter is caused when the most massive galaxy in an SRG is not the most massive one in its host protocluster.


\begin{thebibliography}{110}
\expandafter\ifx\csname natexlab\endcsname\relax\def\natexlab#1{#1}\fi

\bibitem[{{Adams} {et~al.}(2015){Adams}, {Martini}, {Croxall}, {Overzier}, \&
  {Silverman}}]{adams15}
{Adams}, S.~M., {Martini}, P., {Croxall}, K.~V., {Overzier}, R.~A., \&
  {Silverman}, J.~D. 2015, \mnras, 448, 1335

\bibitem[{{{\'A}lvarez Crespo} {et~al.}(2021){{\'A}lvarez Crespo},
  {Smoli{\'c}}, {Finoguenov}, {Barrufet}, \& {Aravena}}]{alvarezcrespo21}
{{\'A}lvarez Crespo}, N., {Smoli{\'c}}, V., {Finoguenov}, A., {Barrufet}, L.,
  \& {Aravena}, M. 2021, \aap, 646, A174

\bibitem[{{Bahcall} \& {West}(1992)}]{bachcall92}
{Bahcall}, N.~A., \& {West}, M.~J. 1992, \apj, 392, 419

\bibitem[{{Bah{\'e}} {et~al.}(2017){Bah{\'e}}, {Barnes}, {Dalla Vecchia},
  {Kay}, {White}, {McCarthy}, {Schaye}, {Bower}, {Crain}, {Theuns}, {Jenkins},
  {McGee}, {Schaller}, {Thomas}, \& {Trayford}}]{bahe17}
{Bah{\'e}}, Y.~M., {Barnes}, D.~J., {Dalla Vecchia}, C., {et~al.} 2017, \mnras,
  470, 4186

\bibitem[{{Bryan} \& {Norman}(1998)}]{bryan98}
{Bryan}, G.~L., \& {Norman}, M.~L. 1998, \apj, 495, 80

\bibitem[{{Cai} {et~al.}(2016){Cai}, {Fan}, {Peirani}, {Bian}, {Frye},
  {McGreer}, {Prochaska}, {Lau}, {Tejos}, {Ho}, \& {Schneider}}]{cai16}
{Cai}, Z., {Fan}, X., {Peirani}, S., {et~al.} 2016, \apj, 833, 135

\bibitem[{{Cai} {et~al.}(2017){Cai}, {Fan}, {Bian}, {Zabludoff}, {Yang},
  {Prochaska}, {McGreer}, {Zheng}, {Kashikawa}, {Wang}, {Frye}, {Green}, \&
  {Jiang}}]{cai17a}
{Cai}, Z., {Fan}, X., {Bian}, F., {et~al.} 2017, \apj, 839, 131

\bibitem[{{Calvi} {et~al.}(2021){Calvi}, {Dannerbauer}, {Arrabal Haro},
  {Rodr{\'\i}guez Espinosa}, {Mu{\~n}oz-Tu{\~n}{\'o}n}, {P{\'e}rez
  Gonz{\'a}lez}, \& {Geier}}]{calvi21}
{Calvi}, R., {Dannerbauer}, H., {Arrabal Haro}, P., {et~al.} 2021, \mnras, 502,
  4558

\bibitem[{{Chabrier}(2003)}]{chabrier03}
{Chabrier}, G. 2003, \apjl, 586, L133

\bibitem[{{Chiang} {et~al.}(2013){Chiang}, {Overzier}, \&
  {Gebhardt}}]{chiang13}
{Chiang}, Y.-K., {Overzier}, R., \& {Gebhardt}, K. 2013, \apj, 779, 127

\bibitem[{{Chiang} {et~al.}(2015){Chiang}, {Overzier}, {Gebhardt},
  {Finkelstein}, {Chiang}, {Hill}, {Blanc}, {Drory}, {Chonis}, {Zeimann},
  {Hagen}, {Schneider}, {Jogee}, {Ciardullo}, \& {Gronwall}}]{chiang15}
{Chiang}, Y.-K., {Overzier}, R.~A., {Gebhardt}, K., {et~al.} 2015, \apj, 808,
  37

\bibitem[{{Choi} \& {Yi}(2017)}]{choi17}
{Choi}, H., \& {Yi}, S.~K. 2017, \apj, 837, 68

\bibitem[{{Cooke} {et~al.}(2014){Cooke}, {Hatch}, {Muldrew}, {Rigby}, \&
  {Kurk}}]{cooke14}
{Cooke}, E.~A., {Hatch}, N.~A., {Muldrew}, S.~I., {Rigby}, E.~E., \& {Kurk},
  J.~D. 2014, \mnras, 440, 3262

\bibitem[{{Cooke} {et~al.}(2019){Cooke}, {Smail}, {Stach}, {Swinbank}, {Bower},
  {Chen}, {Koyama}, \& {Thomson}}]{cooke19}
{Cooke}, E.~A., {Smail}, I., {Stach}, S.~M., {et~al.} 2019, \mnras, 486, 3047

\bibitem[{{Cucciati} {et~al.}(2014){Cucciati}, {Zamorani}, {Lemaux},
  {Bardelli}, {Cimatti}, {Le F{\`e}vre}, {Cassata}, {Garilli}, {Le Brun},
  {Maccagni}, {Pentericci}, {Tasca}, {Thomas}, {Vanzella}, {Zucca}, {Amorin},
  {Capak}, {Cassar{\`a}}, {Castellano}, {Cuby}, {de la Torre}, {Durkalec},
  {Fontana}, {Giavalisco}, {Grazian}, {Hathi}, {Ilbert}, {Moreau}, {Paltani},
  {Ribeiro}, {Salvato}, {Schaerer}, {Scodeggio}, {Sommariva}, {Talia},
  {Taniguchi}, {Tresse}, {Vergani}, {Wang}, {Charlot}, {Contini}, {Fotopoulou},
  {L{\'o}pez-Sanjuan}, {Mellier}, \& {Scoville}}]{cucciati14}
{Cucciati}, O., {Zamorani}, G., {Lemaux}, B.~C., {et~al.} 2014, \aap, 570, A16

\bibitem[{{Cuesta} {et~al.}(2008){Cuesta}, {Prada}, {Klypin}, \&
  {Moles}}]{cuesta08}
{Cuesta}, A.~J., {Prada}, F., {Klypin}, A., \& {Moles}, M. 2008, \mnras, 389,
  385

\bibitem[{{Daddi} {et~al.}(2009){Daddi}, {Dannerbauer}, {Stern}, {Dickinson},
  {Morrison}, {Elbaz}, {Giavalisco}, {Mancini}, {Pope}, \& {Spinrad}}]{daddi09}
{Daddi}, E., {Dannerbauer}, H., {Stern}, D., {et~al.} 2009, \apj, 694, 1517

\bibitem[{{Dalgarno} \& {McCray}(1972)}]{dalgarno72}
{Dalgarno}, A., \& {McCray}, R.~A. 1972, \araa, 10, 375

\bibitem[{{Diener} {et~al.}(2015){Diener}, {Lilly}, {Ledoux}, {Zamorani},
  {Bolzonella}, {Murphy}, {Capak}, {Ilbert}, \& {McCracken}}]{diener15}
{Diener}, C., {Lilly}, S.~J., {Ledoux}, C., {et~al.} 2015, \apj, 802, 31

\bibitem[{{Djorgovski} {et~al.}(2003){Djorgovski}, {Stern}, {Mahabal}, \&
  {Brunner}}]{djorgovski03}
{Djorgovski}, S.~G., {Stern}, D., {Mahabal}, A.~A., \& {Brunner}, R. 2003,
  \apj, 596, 67

\bibitem[{{Dong} {et~al.}(2023){Dong}, {Park}, {Hong}, {Kim}, {Hwang}, {Park},
  \& {Appleby}}]{dong23}
{Dong}, F., {Park}, C., {Hong}, S.~E., {et~al.} 2023, \apj, 953, 98

\bibitem[{{Dubois} {et~al.}(2012){Dubois}, {Devriendt}, {Slyz}, \&
  {Teyssier}}]{dubois12}
{Dubois}, Y., {Devriendt}, J., {Slyz}, A., \& {Teyssier}, R. 2012, \mnras, 420,
  2662

\bibitem[{{Dubois} \& {Teyssier}(2008)}]{dubois08}
{Dubois}, Y., \& {Teyssier}, R. 2008, \aap, 477, 79

\bibitem[{{Dubois} {et~al.}(2014{\natexlab{a}}){Dubois}, {Volonteri}, \&
  {Silk}}]{dubois14a}
{Dubois}, Y., {Volonteri}, M., \& {Silk}, J. 2014{\natexlab{a}}, \mnras, 440,
  1590

\bibitem[{{Dubois} {et~al.}(2014{\natexlab{b}}){Dubois}, {Pichon}, {Welker},
  {Le Borgne}, {Devriendt}, {Laigle}, {Codis}, {Pogosyan}, {Arnouts},
  {Benabed}, {Bertin}, {Blaizot}, {Bouchet}, {Cardoso}, {Colombi}, {de
  Lapparent}, {Desjacques}, {Gavazzi}, {Kassin}, {Kimm}, {McCracken},
  {Milliard}, {Peirani}, {Prunet}, {Rouberol}, {Silk}, {Slyz}, {Sousbie},
  {Teyssier}, {Tresse}, {Treyer}, {Vibert}, \& {Volonteri}}]{dubois14b}
{Dubois}, Y., {Pichon}, C., {Welker}, C., {et~al.} 2014{\natexlab{b}}, \mnras,
  444, 1453

\bibitem[{{Eke} {et~al.}(1996){Eke}, {Cole}, \& {Frenk}}]{eke96}
{Eke}, V.~R., {Cole}, S., \& {Frenk}, C.~S. 1996, \mnras, 282, 263

\bibitem[{{Falco} {et~al.}(2014){Falco}, {Hansen}, {Wojtak}, {Brinckmann},
  {Lindholmer}, \& {Pandolfi}}]{falco14}
{Falco}, M., {Hansen}, S.~H., {Wojtak}, R., {et~al.} 2014, \mnras, 442, 1887

\bibitem[{{Falco} {et~al.}(2013){Falco}, {Mamon}, {Wojtak}, {Hansen}, \&
  {Gottl{\"o}ber}}]{falco13}
{Falco}, M., {Mamon}, G.~A., {Wojtak}, R., {Hansen}, S.~H., \& {Gottl{\"o}ber},
  S. 2013, \mnras, 436, 2639

\bibitem[{{Falder} {et~al.}(2011){Falder}, {Stevens}, {Jarvis}, {Bonfield},
  {Lacy}, {Farrah}, {Oliver}, {Surace}, {Mauduit}, {Vaccari}, {Marchetti},
  {Gonz{\'a}lez-Solares}, {Afonso}, {Cava}, \& {Seymour}}]{falder11}
{Falder}, J.~T., {Stevens}, J.~A., {Jarvis}, M.~J., {et~al.} 2011, \apj, 735,
  123

\bibitem[{{Few} {et~al.}(2012){Few}, {Courty}, {Gibson}, {Kawata}, {Calura}, \&
  {Teyssier}}]{few12}
{Few}, C.~G., {Courty}, S., {Gibson}, B.~K., {et~al.} 2012, \mnras, 424, L11

\bibitem[{{Fu} {et~al.}(2013){Fu}, {Cooray}, {Feruglio}, {Ivison}, {Riechers},
  {Gurwell}, {Bussmann}, {Harris}, {Altieri}, {Aussel}, {Baker}, {Bock},
  {Boylan-Kolchin}, {Bridge}, {Calanog}, {Casey}, {Cava}, {Chapman},
  {Clements}, {Conley}, {Cox}, {Farrah}, {Frayer}, {Hopwood}, {Jia}, {Magdis},
  {Marsden}, {Mart{\'\i}nez-Navajas}, {Negrello}, {Neri}, {Oliver}, {Omont},
  {Page}, {P{\'e}rez-Fournon}, {Schulz}, {Scott}, {Smith}, {Vaccari},
  {Valtchanov}, {Vieira}, {Viero}, {Wang}, {Wardlow}, \& {Zemcov}}]{fu13}
{Fu}, H., {Cooray}, A., {Feruglio}, C., {et~al.} 2013, \nat, 498, 338

\bibitem[{{Greve} {et~al.}(2007){Greve}, {Stern}, {Ivison}, {De Breuck},
  {Kov{\'a}cs}, \& {Bertoldi}}]{greve07}
{Greve}, T.~R., {Stern}, D., {Ivison}, R.~J., {et~al.} 2007, \mnras, 382, 48

\bibitem[{{Gunn} \& {Gott}(1972)}]{gunn72}
{Gunn}, J.~E., \& {Gott}, J.~Richard, I. 1972, \apj, 176, 1

\bibitem[{{Guo} {et~al.}(2011){Guo}, {White}, {Boylan-Kolchin}, {De Lucia},
  {Kauffmann}, {Lemson}, {Li}, {Springel}, \& {Weinmann}}]{guo11}
{Guo}, Q., {White}, S., {Boylan-Kolchin}, M., {et~al.} 2011, \mnras, 413, 101

\bibitem[{{Guzzo} {et~al.}(1997){Guzzo}, {Strauss}, {Fisher}, {Giovanelli}, \&
  {Haynes}}]{guzzo97}
{Guzzo}, L., {Strauss}, M.~A., {Fisher}, K.~B., {Giovanelli}, R., \& {Haynes},
  M.~P. 1997, \apj, 489, 37

\bibitem[{{Haardt} \& {Madau}(1996)}]{haardt96}
{Haardt}, F., \& {Madau}, P. 1996, \apj, 461, 20

\bibitem[{{Hahn} \& {Abel}(2011)}]{hahn11}
{Hahn}, O., \& {Abel}, T. 2011, \mnras, 415, 2101

\bibitem[{{Hamilton}(1998)}]{hamilton98}
{Hamilton}, A.~J.~S. 1998, in Astrophysics and Space Science Library, Vol. 231,
  The Evolving Universe, ed. D.~{Hamilton}, 185

\bibitem[{{Hatch} {et~al.}(2011{\natexlab{a}}){Hatch}, {Kurk}, {Pentericci},
  {Venemans}, {Kuiper}, {Miley}, \& {R{\"o}ttgering}}]{hatch11b}
{Hatch}, N.~A., {Kurk}, J.~D., {Pentericci}, L., {et~al.} 2011{\natexlab{a}},
  \mnras, 415, 2993

\bibitem[{{Hatch} {et~al.}(2011{\natexlab{b}}){Hatch}, {De Breuck}, {Galametz},
  {Miley}, {Overzier}, {R{\"o}ttgering}, {Doherty}, {Kodama}, {Kurk},
  {Seymour}, {Venemans}, {Vernet}, \& {Zirm}}]{hatch11a}
{Hatch}, N.~A., {De Breuck}, C., {Galametz}, A., {et~al.} 2011{\natexlab{b}},
  \mnras, 410, 1537

\bibitem[{{Hayashi} {et~al.}(2012){Hayashi}, {Kodama}, {Tadaki}, {Koyama}, \&
  {Tanaka}}]{hayashi12}
{Hayashi}, M., {Kodama}, T., {Tadaki}, K.-i., {Koyama}, Y., \& {Tanaka}, I.
  2012, \apj, 757, 15

\bibitem[{{Hennawi} {et~al.}(2015){Hennawi}, {Prochaska}, {Cantalupo}, \&
  {Arrigoni-Battaia}}]{hennawi15}
{Hennawi}, J.~F., {Prochaska}, J.~X., {Cantalupo}, S., \& {Arrigoni-Battaia},
  F. 2015, Science, 348, 779

\bibitem[{{Hong} {et~al.}(2016){Hong}, {Park}, \& {Kim}}]{hong16}
{Hong}, S.~E., {Park}, C., \& {Kim}, J. 2016, \apj, 823, 103

\bibitem[{{Husband} {et~al.}(2013){Husband}, {Bremer}, {Stanway}, {Davies},
  {Lehnert}, \& {Douglas}}]{husband13}
{Husband}, K., {Bremer}, M.~N., {Stanway}, E.~R., {et~al.} 2013, \mnras, 432,
  2869

\bibitem[{{Jung} {et~al.}(2014){Jung}, {Lee}, \& {Yi}}]{jung14}
{Jung}, I., {Lee}, J., \& {Yi}, S.~K. 2014, \apj, 794, 74

\bibitem[{{Kaiser}(1984)}]{Kaiser1984}
{Kaiser}, N. 1984, \apjl, 284, L9

\bibitem[{{Kaiser}(1987)}]{kaiser87}
---. 1987, \mnras, 227, 1

\bibitem[{{Kim} {et~al.}(2022){Kim}, {Lee}, {Laigle}, {Dubois}, {Kim}, {Park},
  {Pichon}, {Gibson}, {Few}, {Shin}, \& {Snaith}}]{kim+22}
{Kim}, J., {Lee}, J., {Laigle}, C., {et~al.} 2022, arXiv e-prints,
  arXiv:2212.14539

\bibitem[{{Le Fevre} {et~al.}(1996){Le Fevre}, {Deltorn}, {Crampton}, \&
  {Dickinson}}]{lefevre96}
{Le Fevre}, O., {Deltorn}, J.~M., {Crampton}, D., \& {Dickinson}, M. 1996,
  \apjl, 471, L11

\bibitem[{{Lee} {et~al.}(2014{\natexlab{a}}){Lee}, {Yi}, {Elahi}, {Thomas},
  {Pearce}, {Behroozi}, {Han}, {Helly}, {Jung}, {Knebe}, {Mao}, {Onions},
  {Rodriguez-Gomez}, {Schneider}, {Srisawat}, \& {Tweed}}]{leej14}
{Lee}, J., {Yi}, S.~K., {Elahi}, P.~J., {et~al.} 2014{\natexlab{a}}, \mnras,
  445, 4197

\bibitem[{{Lee} {et~al.}(2021){Lee}, {Shin}, {Snaith}, {Kim}, {Few},
  {Devriendt}, {Dubois}, {Cox}, {Hong}, {Kwon}, {Park}, {Pichon}, {Kim},
  {Gibson}, \& {Park}}]{lee21}
{Lee}, J., {Shin}, J., {Snaith}, O.~N., {et~al.} 2021, \apj, 908, 11

\bibitem[{{Lee} {et~al.}(2014{\natexlab{b}}){Lee}, {Hennawi}, {White}, {Croft},
  \& {Ozbek}}]{lee14a}
{Lee}, K.-G., {Hennawi}, J.~F., {White}, M., {Croft}, R. A.~C., \& {Ozbek}, M.
  2014{\natexlab{b}}, \apj, 788, 49

\bibitem[{{Lee} {et~al.}(2014{\natexlab{c}}){Lee}, {Dey}, {Hong}, {Reddy},
  {Wilson}, {Jannuzi}, {Inami}, \& {Gonzalez}}]{lee14}
{Lee}, K.-S., {Dey}, A., {Hong}, S., {et~al.} 2014{\natexlab{c}}, \apj, 796,
  126

\bibitem[{{Lemaux} {et~al.}(2014){Lemaux}, {Cucciati}, {Tasca}, {Le F{\`e}vre},
  {Zamorani}, {Cassata}, {Garilli}, {Le Brun}, {Maccagni}, {Pentericci},
  {Thomas}, {Vanzella}, {Zucca}, {Amor{\'\i}n}, {Bardelli}, {Capak},
  {Cassar{\`a}}, {Castellano}, {Cimatti}, {Cuby}, {de la Torre}, {Durkalec},
  {Fontana}, {Giavalisco}, {Grazian}, {Hathi}, {Ilbert}, {Moreau}, {Paltani},
  {Ribeiro}, {Salvato}, {Schaerer}, {Scodeggio}, {Sommariva}, {Talia},
  {Taniguchi}, {Tresse}, {Vergani}, {Wang}, {Charlot}, {Contini}, {Fotopoulou},
  {Gal}, {Kocevski}, {L{\'o}pez-Sanjuan}, {Lubin}, {Mellier}, {Sadibekova}, \&
  {Scoville}}]{lemaux14}
{Lemaux}, B.~C., {Cucciati}, O., {Tasca}, L.~A.~M., {et~al.} 2014, \aap, 572,
  A41

\bibitem[{{L'Huillier} {et~al.}(2014){L'Huillier}, {Park}, \&
  {Kim}}]{lhuillier14}
{L'Huillier}, B., {Park}, C., \& {Kim}, J. 2014, \na, 30, 79

\bibitem[{{Lilly} {et~al.}(1999){Lilly}, {Eales}, {Gear}, {Hammer}, {Le
  F{\`e}vre}, {Crampton}, {Bond}, \& {Dunne}}]{lilly99}
{Lilly}, S.~J., {Eales}, S.~A., {Gear}, W. K.~P., {et~al.} 1999, \apj, 518, 641

\bibitem[{{Mamon} {et~al.}(2004){Mamon}, {Sanchis}, {Salvador-Sol{\'e}}, \&
  {Solanes}}]{mamon04}
{Mamon}, G.~A., {Sanchis}, T., {Salvador-Sol{\'e}}, E., \& {Solanes}, J.~M.
  2004, \aap, 414, 445

\bibitem[{{McConachie} {et~al.}(2022){McConachie}, {Wilson}, {Forrest},
  {Marsan}, {Muzzin}, {Cooper}, {Annunziatella}, {Marchesini}, {Chan}, {Gomez},
  {Abdullah}, {Saracco}, \& {Nantais}}]{mcconachie22}
{McConachie}, I., {Wilson}, G., {Forrest}, B., {et~al.} 2022, \apj, 926, 37

\bibitem[{{Muldrew} {et~al.}(2015){Muldrew}, {Hatch}, \& {Cooke}}]{muldrew15}
{Muldrew}, S.~I., {Hatch}, N.~A., \& {Cooke}, E.~A. 2015, \mnras, 452, 2528

\bibitem[{{Muldrew} {et~al.}(2018){Muldrew}, {Hatch}, \& {Cooke}}]{muldrew18}
---. 2018, \mnras, 473, 2335

\bibitem[{{Newman} {et~al.}(2022){Newman}, {Rudie}, {Blanc}, {Qezlou}, {Bird},
  {Kelson}, {P{\'e}rez}, {Congiu}, {Lemaux}, {Dressler}, \&
  {Mulchaey}}]{newman22}
{Newman}, A.~B., {Rudie}, G.~C., {Blanc}, G.~A., {et~al.} 2022, \nat, 606, 475

\bibitem[{{O{\~n}orbe} {et~al.}(2014){O{\~n}orbe}, {Garrison-Kimmel}, {Maller},
  {Bullock}, {Rocha}, \& {Hahn}}]{onorbe+14}
{O{\~n}orbe}, J., {Garrison-Kimmel}, S., {Maller}, A.~H., {et~al.} 2014,
  \mnras, 437, 1894

\bibitem[{{Oteo} {et~al.}(2018){Oteo}, {Ivison}, {Dunne}, {Manilla-Robles},
  {Maddox}, {Lewis}, {de Zotti}, {Bremer}, {Clements}, {Cooray}, {Dannerbauer},
  {Eales}, {Greenslade}, {Omont}, {Perez{\textendash}Fourn{\'o}n}, {Riechers},
  {Scott}, {van der Werf}, {Weiss}, \& {Zhang}}]{oteo18}
{Oteo}, I., {Ivison}, R.~J., {Dunne}, L., {et~al.} 2018, \apj, 856, 72

\bibitem[{{Ouchi} {et~al.}(2005){Ouchi}, {Shimasaku}, {Akiyama}, {Sekiguchi},
  {Furusawa}, {Okamura}, {Kashikawa}, {Iye}, {Kodama}, {Saito}, {Sasaki},
  {Simpson}, {Takata}, {Yamada}, {Yamanoi}, {Yoshida}, \& {Yoshida}}]{ouchi05}
{Ouchi}, M., {Shimasaku}, K., {Akiyama}, M., {et~al.} 2005, \apjl, 620, L1

\bibitem[{{Overzier}(2016)}]{overzier16}
{Overzier}, R.~A. 2016, \aapr, 24, 14

\bibitem[{{Pace} {et~al.}(2010){Pace}, {Waizmann}, \& {Bartelmann}}]{pace10}
{Pace}, F., {Waizmann}, J.~C., \& {Bartelmann}, M. 2010, \mnras, 406, 1865

\bibitem[{{Park} {et~al.}(2022){Park}, {Lee}, {Kim}, {Jeong}, {Pichon},
  {Gibson}, {Snaith}, {Shin}, {Kim}, {Dubois}, \& {Few}}]{park22}
{Park}, C., {Lee}, J., {Kim}, J., {et~al.} 2022, \apj, 937, 15

\bibitem[{{Pascarelle} {et~al.}(1996){Pascarelle}, {Windhorst}, {Keel}, \&
  {Odewahn}}]{pascarelle96}
{Pascarelle}, S.~M., {Windhorst}, R.~A., {Keel}, W.~C., \& {Odewahn}, S.~C.
  1996, \nat, 383, 45

\bibitem[{{Peebles}(1980)}]{peebles80}
{Peebles}, P.~J.~E. 1980, {The large-scale structure of the universe}

\bibitem[{{Peebles}(1984)}]{peebles84}
---. 1984, \apj, 284, 439

\bibitem[{{Planck Collaboration} {et~al.}(2016){Planck Collaboration}, {Ade},
  \& {Aghanim}}]{planck16}
{Planck Collaboration}, {Ade}, P.~A.~R., \& {Aghanim}, N. e.~a. 2016, \aap,
  594, A13

\bibitem[{{Prescott} {et~al.}(2008){Prescott}, {Kashikawa}, {Dey}, \&
  {Matsuda}}]{prescott08}
{Prescott}, M. K.~M., {Kashikawa}, N., {Dey}, A., \& {Matsuda}, Y. 2008, \apjl,
  678, L77

\bibitem[{{Prescott} {et~al.}(2012){Prescott}, {Dey}, {Brodwin}, {Chaffee},
  {Desai}, {Eisenhardt}, {Le Floc'h}, {Jannuzi}, {Kashikawa}, {Matsuda}, \&
  {Soifer}}]{prescott12}
{Prescott}, M. K.~M., {Dey}, A., {Brodwin}, M., {et~al.} 2012, \apj, 752, 86

\bibitem[{{Ramakrishnan} {et~al.}(2022){Ramakrishnan}, {Moon}, {Dey}, {Farooq},
  {Gawiser}, {Huang}, {Lee}, {Park}, {Valdes}, \& {Yang}}]{ramakrishnan22}
{Ramakrishnan}, V., {Moon}, B., {Dey}, A., {et~al.} 2022, in American
  Astronomical Society Meeting Abstracts, Vol.~54, American Astronomical
  Society Meeting Abstracts, 428.05

\bibitem[{{Rasera} \& {Teyssier}(2006)}]{rasera06}
{Rasera}, Y., \& {Teyssier}, R. 2006, \aap, 445, 1

\bibitem[{{Rines} \& {Diaferio}(2006)}]{rines06}
{Rines}, K., \& {Diaferio}, A. 2006, \aj, 132, 1275

\bibitem[{{Rotermund} {et~al.}(2021){Rotermund}, {Chapman}, {Phadke}, {Hill},
  {Pass}, {Aravena}, {Ashby}, {Babul}, {B{\'e}thermin}, {Canning}, {de Breuck},
  {Dong}, {Gonzalez}, {Hayward}, {Jarugula}, {Marrone}, {Narayanan}, {Reuter},
  {Scott}, {Spilker}, {Vieira}, {Wang}, \& {Weiss}}]{rotermund21}
{Rotermund}, K.~M., {Chapman}, S.~C., {Phadke}, K.~A., {et~al.} 2021, \mnras,
  502, 1797

\bibitem[{{Scoccimarro}(1998)}]{scoccimarro98}
{Scoccimarro}, R. 1998, \mnras, 299, 1097

\bibitem[{{Shen} {et~al.}(2021){Shen}, {Lemaux}, {Lubin}, {Cucciati}, {Le
  F{\`e}vre}, {Liu}, {Fang}, {Pelliccia}, {Tomczak}, {McKean}, {Miller},
  {Fassnacht}, {Gal}, {Hung}, {Hathi}, {Bardelli}, {Vergani}, \&
  {Zucca}}]{shen21}
{Shen}, L., {Lemaux}, B.~C., {Lubin}, L.~M., {et~al.} 2021, \apj, 912, 60

\bibitem[{{Shi} {et~al.}(2019){Shi}, {Huang}, {Lee}, {Toshikawa}, {Bowen},
  {Malavasi}, {Lemaux}, {Cucciati}, {Le Fevre}, \& {Dey}}]{shi19}
{Shi}, K., {Huang}, Y., {Lee}, K.-S., {et~al.} 2019, \apj, 879, 9

\bibitem[{{Shimasaku} {et~al.}(2003){Shimasaku}, {Ouchi}, {Okamura},
  {Kashikawa}, {Doi}, {Furusawa}, {Hamabe}, {Hayashino}, {Kawabata}, {Kimura},
  {Kodaira}, {Komiyama}, {Matsuda}, {Miyazaki}, {Miyazaki}, {Nakata}, {Ohta},
  {Ohyama}, {Sekiguchi}, {Shioya}, {Tamura}, {Taniguchi}, {Yagi}, {Yamada}, \&
  {Yasuda}}]{shimasaku03}
{Shimasaku}, K., {Ouchi}, M., {Okamura}, S., {et~al.} 2003, \apjl, 586, L111

\bibitem[{{Springel} {et~al.}(2005){Springel}, {White}, {Jenkins}, {Frenk},
  {Yoshida}, {Gao}, {Navarro}, {Thacker}, {Croton}, {Helly}, {Peacock}, {Cole},
  {Thomas}, {Couchman}, {Evrard}, {Colberg}, \& {Pearce}}]{springel05}
{Springel}, V., {White}, S. D.~M., {Jenkins}, A., {et~al.} 2005, \nat, 435, 629

\bibitem[{{Stark} {et~al.}(2015){Stark}, {White}, {Lee}, \&
  {Hennawi}}]{stark15}
{Stark}, C.~W., {White}, M., {Lee}, K.-G., \& {Hennawi}, J.~F. 2015, \mnras,
  453, 311

\bibitem[{{Steidel} {et~al.}(1998){Steidel}, {Adelberger}, {Dickinson},
  {Giavalisco}, {Pettini}, \& {Kellogg}}]{steidel98}
{Steidel}, C.~C., {Adelberger}, K.~L., {Dickinson}, M., {et~al.} 1998, \apj,
  492, 428

\bibitem[{{Steidel} {et~al.}(2005){Steidel}, {Adelberger}, {Shapley}, {Erb},
  {Reddy}, \& {Pettini}}]{steidel05}
{Steidel}, C.~C., {Adelberger}, K.~L., {Shapley}, A.~E., {et~al.} 2005, \apj,
  626, 44

\bibitem[{{Steidel} {et~al.}(2000){Steidel}, {Adelberger}, {Shapley},
  {Pettini}, {Dickinson}, \& {Giavalisco}}]{steidel00}
---. 2000, \apj, 532, 170

\bibitem[{{Stevens} {et~al.}(2010){Stevens}, {Jarvis}, {Coppin}, {Page},
  {Greve}, {Carrera}, \& {Ivison}}]{stevens10}
{Stevens}, J.~A., {Jarvis}, M.~J., {Coppin}, K.~E.~K., {et~al.} 2010, \mnras,
  405, 2623

\bibitem[{{Stevens} {et~al.}(2003){Stevens}, {Ivison}, {Dunlop}, {Smail},
  {Percival}, {Hughes}, {R{\"o}ttgering}, {van Breugel}, \&
  {Reuland}}]{stevens03}
{Stevens}, J.~A., {Ivison}, R.~J., {Dunlop}, J.~S., {et~al.} 2003, \nat, 425,
  264

\bibitem[{{Sutherland} \& {Dopita}(1993)}]{sutherland93}
{Sutherland}, R.~S., \& {Dopita}, M.~A. 1993, \apjs, 88, 253

\bibitem[{{Suto} {et~al.}(2016){Suto}, {Kitayama}, {Osato}, {Sasaki}, \&
  {Suto}}]{suto16}
{Suto}, D., {Kitayama}, T., {Osato}, K., {Sasaki}, S., \& {Suto}, Y. 2016,
  \pasj, 68, 14

\bibitem[{{Teyssier}(2002)}]{teyssier02}
{Teyssier}, R. 2002, \aap, 385, 337

\bibitem[{{Toft} {et~al.}(2014){Toft}, {Smol{\v{c}}i{\'c}}, {Magnelli},
  {Karim}, {Zirm}, {Michalowski}, {Capak}, {Sheth}, {Schawinski}, {Krogager},
  {Wuyts}, {Sanders}, {Man}, {Lutz}, {Staguhn}, {Berta}, {Mccracken}, {Krpan},
  \& {Riechers}}]{toft14}
{Toft}, S., {Smol{\v{c}}i{\'c}}, V., {Magnelli}, B., {et~al.} 2014, \apj, 782,
  68

\bibitem[{{Toshikawa} {et~al.}(2012){Toshikawa}, {Kashikawa}, {Ota},
  {Morokuma}, {Shibuya}, {Hayashi}, {Nagao}, {Jiang}, {Malkan}, {Egami},
  {Shimasaku}, {Motohara}, \& {Ishizaki}}]{toshikawa12}
{Toshikawa}, J., {Kashikawa}, N., {Ota}, K., {et~al.} 2012, \apj, 750, 137

\bibitem[{{Toshikawa} {et~al.}(2014){Toshikawa}, {Kashikawa}, {Overzier},
  {Shibuya}, {Ishikawa}, {Ota}, {Shimasaku}, {Tanaka}, {Hayashi}, {Niino}, \&
  {Onoue}}]{toshikawa14}
{Toshikawa}, J., {Kashikawa}, N., {Overzier}, R., {et~al.} 2014, \apj, 792, 15

\bibitem[{{Toshikawa} {et~al.}(2016){Toshikawa}, {Kashikawa}, {Overzier},
  {Malkan}, {Furusawa}, {Ishikawa}, {Onoue}, {Ota}, {Tanaka}, {Niino}, \&
  {Uchiyama}}]{toshikawa16}
---. 2016, \apj, 826, 114

\bibitem[{{Toshikawa} {et~al.}(2018){Toshikawa}, {Uchiyama}, {Kashikawa},
  {Ouchi}, {Overzier}, {Ono}, {Harikane}, {Ishikawa}, {Kodama}, {Matsuda},
  {Lin}, {Onoue}, {Tanaka}, {Nagao}, {Akiyama}, {Komiyama}, {Goto}, \&
  {Lee}}]{toshikawa18}
{Toshikawa}, J., {Uchiyama}, H., {Kashikawa}, N., {et~al.} 2018, \pasj, 70, S12

\bibitem[{{Trebitsch} {et~al.}(2021){Trebitsch}, {Dubois}, {Volonteri},
  {Pfister}, {Cadiou}, {Katz}, {Rosdahl}, {Kimm}, {Pichon}, {Beckmann},
  {Devriendt}, \& {Slyz}}]{Trebitsch2021}
{Trebitsch}, M., {Dubois}, Y., {Volonteri}, M., {et~al.} 2021, \aap, 653, A154

\bibitem[{{Truong} {et~al.}(2018){Truong}, {Rasia}, {Mazzotta}, {Planelles},
  {Biffi}, {Fabjan}, {Beck}, {Borgani}, {Dolag}, {Gaspari}, {Granato},
  {Murante}, {Ragone-Figueroa}, \& {Steinborn}}]{truong18}
{Truong}, N., {Rasia}, E., {Mazzotta}, P., {et~al.} 2018, \mnras, 474, 4089

\bibitem[{{Umehata} {et~al.}(2014){Umehata}, {Tamura}, {Kohno}, {Hatsukade},
  {Scott}, {Kubo}, {Yamada}, {Ivison}, {Cybulski}, {Aretxaga}, {Austermann},
  {Hughes}, {Ezawa}, {Hayashino}, {Ikarashi}, {Iono}, {Kawabe}, {Matsuda},
  {Matsuo}, {Nakanishi}, {Oshima}, {Perera}, {Takata}, {Wilson}, \&
  {Yun}}]{umehata14}
{Umehata}, H., {Tamura}, Y., {Kohno}, K., {et~al.} 2014, \mnras, 440, 3462

\bibitem[{{Umehata} {et~al.}(2015){Umehata}, {Tamura}, {Kohno}, {Ivison},
  {Alexander}, {Geach}, {Hatsukade}, {Hughes}, {Ikarashi}, {Kato}, {Izumi},
  {Kawabe}, {Kubo}, {Lee}, {Lehmer}, {Makiya}, {Matsuda}, {Nakanishi}, {Saito},
  {Smail}, {Yamada}, {Yamaguchi}, \& {Yun}}]{umehata15}
---. 2015, \apjl, 815, L8

\bibitem[{{Venemans} {et~al.}(2002){Venemans}, {Kurk}, {Miley},
  {R{\"o}ttgering}, {van Breugel}, {Carilli}, {De Breuck}, {Ford}, {Heckman},
  {McCarthy}, \& {Pentericci}}]{venemans02}
{Venemans}, B.~P., {Kurk}, J.~D., {Miley}, G.~K., {et~al.} 2002, \apjl, 569,
  L11

\bibitem[{{Venemans} {et~al.}(2004){Venemans}, {R{\"o}ttgering}, {Overzier},
  {Miley}, {De Breuck}, {Kurk}, {van Breugel}, {Carilli}, {Ford}, {Heckman},
  {McCarthy}, \& {Pentericci}}]{venemans04}
{Venemans}, B.~P., {R{\"o}ttgering}, H.~J.~A., {Overzier}, R.~A., {et~al.}
  2004, \aap, 424, L17

\bibitem[{{Venemans} {et~al.}(2005){Venemans}, {R{\"o}ttgering}, {Miley},
  {Kurk}, {De Breuck}, {Overzier}, {van Breugel}, {Carilli}, {Ford}, {Heckman},
  {Pentericci}, \& {McCarthy}}]{venemans05}
{Venemans}, B.~P., {R{\"o}ttgering}, H.~J.~A., {Miley}, G.~K., {et~al.} 2005,
  \aap, 431, 793

\bibitem[{{Venemans} {et~al.}(2007){Venemans}, {R{\"o}ttgering}, {Miley}, {van
  Breugel}, {de Breuck}, {Kurk}, {Pentericci}, {Stanford}, {Overzier}, {Croft},
  \& {Ford}}]{venemans07}
---. 2007, \aap, 461, 823

\bibitem[{{Wang} {et~al.}(2021){Wang}, {Mo}, {Li}, \& {Chen}}]{wang21}
{Wang}, K., {Mo}, H.~J., {Li}, C., \& {Chen}, Y. 2021, \mnras, 505, 3892

\bibitem[{{Wang} {et~al.}(2016){Wang}, {Elbaz}, {Daddi}, {Finoguenov}, {Liu},
  {Schreiber}, {Mart{\'\i}n}, {Strazzullo}, {Valentino}, {van der Burg},
  {Zanella}, {Ciesla}, {Gobat}, {Le Brun}, {Pannella}, {Sargent}, {Shu}, {Tan},
  {Cappelluti}, \& {Li}}]{wang16}
{Wang}, T., {Elbaz}, D., {Daddi}, E., {et~al.} 2016, \apj, 828, 56

\bibitem[{{Wojtak} {et~al.}(2005){Wojtak}, {{\L}okas}, {Gottl{\"o}ber}, \&
  {Mamon}}]{wojtak05}
{Wojtak}, R., {{\L}okas}, E.~L., {Gottl{\"o}ber}, S., \& {Mamon}, G.~A. 2005,
  \mnras, 361, L1

\bibitem[{{Wold} {et~al.}(2003){Wold}, {Armus}, {Neugebauer}, {Jarrett}, \&
  {Lehnert}}]{wold03}
{Wold}, M., {Armus}, L., {Neugebauer}, G., {Jarrett}, T.~H., \& {Lehnert},
  M.~D. 2003, \aj, 126, 1776

\bibitem[{{Yajima} {et~al.}(2022){Yajima}, {Abe}, {Khochfar}, {Nagamine},
  {Inoue}, {Kodama}, {Arata}, {Dalla Vecchia}, {Fukushima}, {Hashimoto},
  {Kashikawa}, {Kubo}, {Li}, {Matsuda}, {Mawatari}, {Ouchi}, \&
  {Umehata}}]{yajima22}
{Yajima}, H., {Abe}, M., {Khochfar}, S., {et~al.} 2022, \mnras, 509, 4037

\bibitem[{{Yonekura} {et~al.}(2022){Yonekura}, {Kajisawa}, {Hamaguchi},
  {Mawatari}, \& {Yamada}}]{yonekura22}
{Yonekura}, N., {Kajisawa}, M., {Hamaguchi}, E., {Mawatari}, K., \& {Yamada},
  T. 2022, \apj, 930, 102

\end{thebibliography}
\end{document}